\documentclass{IEEEtran}
\usepackage{cite}
\usepackage{amsmath,amssymb,amsfonts}
\usepackage{algorithmic}
\usepackage{graphicx}
\usepackage{textcomp}
\usepackage{siunitx}
\usepackage{tikz}
\usepackage{nicematrix}

\usetikzlibrary{math, calc}
\usetikzlibrary{angles} 
\usetikzlibrary{shapes, arrows.meta, positioning}

\newcommand{\myVec}[1]{\boldsymbol{#1}}

\def\BibTeX{{\rm B\kern-.05em{\sc i\kern-.025em b}\kern-.08em
    T\kern-.1667em\lower.7ex\hbox{E}\kern-.125emX}}
\begin{document}
\title{Optimizing Sparse Antenna Arrays for Localization and Sensing using Vector Spherical Wave Functions}
\author{Tobias Lafer, Erik Leitinger, and Klaus Witrisal
\thanks{Manuscript received 18 September 2025. The work presented in this document was conducted in the frame of the SINFONIA project. The SINFONIA has received funding from the Recovery and Resilience Facility (RRF) as the centrepiece of NextGenerationEU via the Austrian Research Promotion Agency (FFG) and Austria Wirtschaftsservice Gesellschaft mbH (aws) in the frame of the IPCEI ME/CT – Important Project of Common European Interest on Microelectronic and Communication Technologies under FFG project No 917423 and AWS project No P2431566.}
\thanks{Tobias Lafer is with NXP Semiconductors Austria, Gratkorn 8101, Austria and Graz University of Technology, Graz 8010, Austria (e-mail: tobiasflorian.lafer@nxp.com).}
\thanks{Erik Leitinger and Klaus Witrisal are with Graz University of Technology, Graz 8010, Austria (e-mail: witrisal@tugraz.at).}
}

\maketitle

\begin{abstract} 
    In increasing number of electronic devices implement wideband radio technologies for localization and sensing purposes, like ultra-wideband (UWB). Such radio technologies benefit from a large number of antennas, but space for antennas is often limited, especially in devices for mobile and IoT applications. A common challenge is therefore to optimize the placement and orientations of a small number of antenna elements inside a device, leading to the best localization performance. We propose a method for systematically approaching the optimization of such sparse arrays by means of Cram\'er-Rao lower bounds (CRLBs) and vector spherical wave functions (VSWFs).
    The VSWFs form the basis of a wideband signal model considering frequency, direction and polarization-dependent characteristics of the antenna array under test (AUT), together with mutual coupling and distortions from surrounding obstacles. We derive the CRLBs for localization parameters like delay and angle-of-arrival for this model under additive white Gaussian noise channel conditions, and formulate optimization problems for determining optimal antenna positions and orientations via minimization of the CRLBs.
    The proposed optimization procedure is demonstrated by means of an exemplary arrangement of three \textit{Crossed Exponentially Tapered Slot (XETS)} antennas.
\end{abstract}

\bstctlcite{IEEEexample:BSTcontrol}

\section{Introduction}
\label{sec:introduction}

\begin{figure}
    \centering
    \includegraphics[width=\columnwidth]{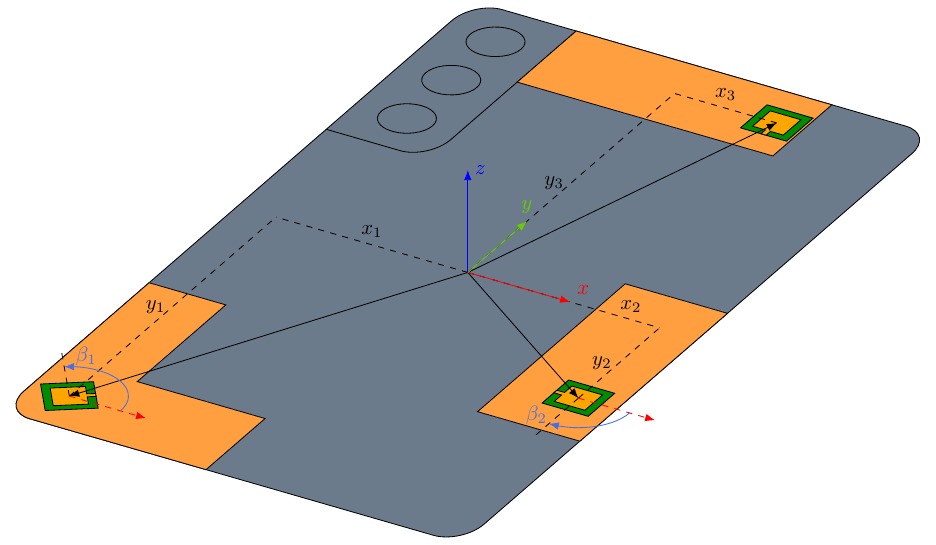}
    \caption{An example antenna positioning problem. The positions $x_k$ and $y_k$, as well as the orientations $\beta_k$ of three antenna elements within the orange-shaded areas shall be determined, such that the localization performance is maximized.}
    \label{fig:antenna_positioning_problem}
\end{figure}
Ultra-Wideband (UWB) is a well-established radio technology for low-power wireless localization in indoor environments \cite{coppensOverviewUWBStandards2022a}. It has therefore become widely adopted in mobile and IoT devices, where the available space for antenna elements is often severely constrained. Consequently, only a small number of antenna elements can typically be integrated, despite the fact that the localization accuracy generally improves with an increasing number of antennas \cite{wildingAccuracyBoundsArrayBased2018, hanPerformanceLimitsGeometric2016}. 
Moreover, the antenna performance in such compact devices is affected by electromagnetic coupling to nearby components, for example the camera assembly in a smartphone. A key challenge in the integration of UWB into mobile and IoT platforms is thus to determine the placement and orientation of a limited number of antenna elements that maximize the achievable localization accuracy. An illustrative example is shown in Figure \ref{fig:antenna_positioning_problem}, where the positions and orientations of three antenna elements inside a smartphone are to be optimized. 
The "localization accuracy", as target for the optimization of the antenna arrangement, is thereby always device and application specific. For example, typical application scenarios of UWB include (i) estimating the position of a single unknown transmitter relative to the agent, or (ii) estimating an agent's position relative to a set of fixed anchors \cite{coppensOverviewUWBStandards2022a}. In both cases, accurate extraction of the delay and direction-of-arrival (DoA) of the line-of-sight (LOS) signal between transmitter and the agent is essential. Since UWB is primarily employed in indoor environments, the corresponding wireless channels are, however, typically characterized by pronounced multipath propagation \cite{molischIEEE802154aChannel2004}.

In antenna position optimization problems, such as the one illustrated in Figure\ref{fig:antenna_positioning_problem}, cost functions can be defined in terms of statistical performance bounds on the achievable localization accuracy of a given antenna configuration, such as the Cram\'er–Rao lower bound (CRLB) \cite{gazzahCramerRaoBoundsAntenna2006, wangAdaptiveArrayThinning2015, keskinOptimumVolumetricArray2020}. The CRLB, obtained as the inverse of the Fisher information matrix (FIM), provides a lower bound on the covariance matrix of any unbiased estimator for a parameter vector \cite{SheWin:TIT2010part1,LeitingerJSAC2015,hanPerformanceLimitsGeometric2016}. In the context of radio positioning, it therefore represents a fundamental limit on the variance of the position error. However, existing derivations of the CRLB for positioning accuracy typically neglect direction- and polarization-dependent antenna characteristics as well as mutual coupling effects \cite{gazzahCramerRaoBoundsAntenna2006, wangAdaptiveArrayThinning2015, keskinOptimumVolumetricArray2020}. Even when such effects are included, variations of antenna characteristics with frequency are usually disregarded \cite{costaDoAPolarizationEstimation2012, costaUnifiedArrayManifold2010, costaSteeringVectorModeling2010, nordeboFundamentalLimitationsDOA2006}. Accounting for all these effects is crucial for the design of UWB antenna arrays. Nevertheless, assumptions of isotropic radiation and frequency independence are still commonly made \cite{malikUltrawidebandAntennaDistortion2008, veitImpactUWBAntennas2020, laferEffectiveHeightAnalysisUWB2025}.
	
A rigorous derivation of the FIM on the positioning error that accounts for direction-, polarization-, and frequency-dependent antenna characteristics requires accurate mathematical models of how antennas shape the received signals. Based on a suitable signal model, the likelihood function of the received signals must first be established \cite{kayFundamentalsStatisticalSignal1993a}, followed by the computation of its partial derivatives with respect to the parameters of interest. In radio signal-based localization, these parameters typically include propagation delay, direction-of-arrival (DoA), direction-of-departure (DoD), Doppler, or polarization. Although the calculation of these derivatives may initially appear cumbersome, the problem simplifies once antenna characteristics are expressed through appropriate spatial basis functions. Common representations include scalar spherical wave functions (SSWFs) \cite{5068319} and vector spherical wave functions (VSWFs) \cite{hansenSphericalNearfieldAntenna1988a}. FIM formulations based on SSWFs, with applications in localization, are presented in \cite{costaDoAPolarizationEstimation2012,costaUnifiedArrayManifold2010,costaSteeringVectorModeling2010}, while VSWF-based analyses are discussed in \cite{nordeboFundamentalLimitationsDOA2006}. All of these derivations assume frequency-independent antenna behavior. VSWFs, in particular, are widely adopted in the antenna community, with applications in full-sphere antenna measurements \cite{soklicFullSphereAntennaMeasurements2024}, characterization of MIMO arrays \cite{ximenesCapacityEvaluationMIMO2010, gustafssonCharacterizationMIMOAntennas2006}, antenna optimization \cite{berkelmannAntennaOptimizationWBAN2022}, and efficient modeling of mutual coupling effects \cite{marinovicFastCharacterizationMutually2021,rubioGeneralizedscatteringmatrixAnalysisClass2005}. VSWFs have also been employed for UWB antenna modeling \cite{5068315,roblinUltraCompressedParametric2006a,jinxinduParametricModelingDeformable2016}.

\subsection{Contributions}

In this paper, we present a systematic framework for optimizing the placement of antennas within UWB antenna arrays. The framework utilizes a cost function based on the Cram\'er-Rao lower bounds (CRLBs) for radio positioning, where the CRLBs explicitly consider direction-, polarization-, and frequency-dependent antenna characteristics. For this purpose, we extend existing models of antenna influences on received signals, such as those in \cite{costaDoAPolarizationEstimation2012,nordeboFundamentalLimitationsDOA2006}, by expressing the reception characteristics of the antenna array under test (AUT) in terms of vector spherical wave functions (VSWFs). This representation naturally captures frequency-dependent effects and implicitly accounts for mutual coupling between antenna elements as well as interactions with nearby objects. Building on this, we calculate the FIM for the propagation delay, direction-of-arrival (DoA), and polarization of a electromagnetic plane wave, incident on the antenna array to be optimized. We thereby restrict the derivations to additive white Gaussian noise (AWGN) channel conditions, as the complexity of common signal models for multipath propagation in indoor environments \cite{SheWin:TIT2010part1,LeitingerJSAC2015,wildingSingleAnchorMultipathAssistedIndoor2018} prevented us from derive a closed form solution for the FIM considering all direction-, polarization-, and frequency-dependent reception characteristics of the antenna array to be optimized. We intend to close this gap in future.
Finally, we formulate two optimization problems that use the FIM. One problem minimizes the trace of the inverse FIM, while the other minimizes its determinant. Both formulations aim at maximizing achievable localization accuracy, but they emphasize different properties of the estimation problem \cite{pukelsheimOptimalDesignExperiments2006a}. By solving these optimization problems iteratively, the positions and orientations of the antenna elements in the AUT are adjusted such that the resulting array configuration optimally supports high-accuracy localization.
The main contributions of this work are summarized as follows.
\begin{itemize}
	\item We introduce a generalized signal model that incorporates direction-, polarization-, and frequency-dependent antenna characteristics using VSWFs.  
	\item We derive the FIM for delay, DoA, and polarization under an AWGN channel model.
	\item We formulate two FIM-based optimization problems for antenna positioning that maximize localization accuracy by minimizing either the trace or the determinant of the inverse FIM.  
\end{itemize}
The rest of this paper is structured as follow: The derivation of the mathematical model for the signals at the antenna ports is covered in Section \ref{sec:wideband_signal_model}. The determination of the AUT reception characteristics by means of VSWFs from a full-wave EM simulation is presented in Section \ref{sec:SWCs_from_simulation}. Section \ref{sec:FIM_and_CRLBs} contains the derivations of the FIM and the CRLBs. The definitions of the two optimization problems are described in Section \ref{sec:optimization_problems}. The optimization procedure is demonstrated by means of a 3-element UWB antenna array in ´Section \ref{sec:example}. The outcomes of this optimization are discussed in Section \ref{sec:example_results}, followed by a wrap-up of the paper together with an outlook on further research in Section \ref{sec:conclusion_and_outlook}.

\section{Wideband Signal Model using VSWFs}
\label{sec:wideband_signal_model}
The total electric field strength $\myVec{E}\in \mathbb{C}^3$ outside a sphere circumscribing an antenna array can be expanded in terms of vector spherical wave functions \cite{hansenSphericalNearfieldAntenna1988a}
\begin{align}
    \myVec{E}(r, \theta, \varphi) = \frac{k}{\sqrt{\eta}} \sum\limits_{j=1}^J a_j \myVec{F}_j^{(4)} \left(kr, \theta, \varphi\right) + b_j \myVec{F}_j^{(3)} \left(kr, \theta, \varphi\right)\ , \label{eq:VSWE_general}
\end{align}
where $\myVec{F}_j^{(3)}\in\mathbb{C}^3$ represent outward propagating VSWFs, and $\myVec{F}_j^{(4)}\in\mathbb{C}^3$ represent inward propagating VSWFs. Radius $r$, azimuth $\varphi$ and elevation $\theta$ are spherical coordinates representing a position in three-dimensional space relative to the center of the circumscribing sphere. Throughout this paper, the spherical coordinate system defined in \cite[Sec. 7]{9714428} is used.
The wavenumber $k$ is given by $k=\frac{\omega}{c_0}$ with $\omega$ being the angular frequency in $\si{\radian\per\second}$ and $c_0$ being the free-space propagation velocity.
The summation over $j$ in \eqref{eq:VSWE_general} is, strictly speaking, an infinite sum, but can be truncated after $J$ terms for the vast majority of cases without significant loss of accuracy \cite{jensenNumberModesSpherical2004}.
Based on the vector spherical wave expansion (VSWE) \eqref{eq:VSWE_general}, the concept of a generalized scattering matrix (GSM) of an antenna or array can be established \cite{hansenSphericalNearfieldAntenna1988a,657097a1-1c9b-4533-b35f-556d739543ff}. Assuming an antenna array with $L$ elements, the GSM $\mathcal{S}\in\mathbb{C}^{JL\times JL}$ describes the linear relationship between the spherical wave coefficients (SWCs) $b_j\in\mathbb{C}$ and $a_j\in\mathbb{C}$ in \eqref{eq:VSWE_general}, and the incident and reflected voltages waves $v^{(l)}\in\mathbb{C}$ and $w^{(l)}\in\mathbb{C}$ at the antenna ports by means of a matrix equation
\begin{align}
    \begin{bmatrix}
        \myVec{b} \\ \myVec{w}
    \end{bmatrix} = 
    \underbrace{\begin{bmatrix}
        \myVec{S} & \myVec{T} \\
        \myVec{R} & \myVec{\Gamma}
    \end{bmatrix}}_{\mathcal{S}}
    \begin{bmatrix}
        \myVec{a} \\ \myVec{v}
    \end{bmatrix} \label{eq:GSM_antenna_arrays}\ ,
\end{align}
using the vector representations of SWCs and port voltages
\begin{align*}
    \myVec{b} := \left[b_1, \hdots, b_J \right]^T&,\ \myVec{w}:= \left[w^{(1)}, \hdots, w^{(L)}\right]^T, \\
    \myVec{a} := \left[a_1, \hdots, a_J \right]^T&,\ \myVec{v}:= \left[v^{(1)}, \hdots, v^{(L)}\right]^T.
\end{align*}
In \eqref{eq:GSM_antenna_arrays}, the antenna scattering matrix $\myVec{S}\in\mathbb{C}^{J \times J}$ connects inward and outward propagating spherical modes $a_j$ and $b_j$, and thus describes the electromagnetic scattering properties of the array structure. Matrix $\myVec{T}\in\mathbb{C}^{J\times L}$ contains the antenna transmission coefficients $T_{j}^{(l)}$, and describes conversion of the incident port voltage waves $v^{(l)}$ to the outward propagating spherical modes $b_j$. Matrix $\myVec{R}\in\mathbb{C}^{L\times J}$ contains the antenna reception coefficients $R_{j}^{(l)}$, and describes the conversion of incident spherical modes $a_j$ to outward propagating voltages waves $w^{(l)}$ at the antenna ports. The remaining matrix $\myVec{\Gamma}\in\mathbb{C}^{L \times L}$ is a standard circuit-level scattering matrix. 
The summation over mode index $j$ in \eqref{eq:VSWE_general} is actually a triple-sum over three indices $s$, $m$ and $n$, which have been merged into a single index by
\begin{align}
    j := 2\left[n(n+1) + m-1\right] &+ s,\ J:=2N(N+2) \label{eq:single_index_from_triple_index}\\
    \sum\limits_{s=1}^{2} \sum\limits_{m=-n}^m \sum\limits_{n=1}^N &\rightarrow \sum\limits_{j=1}^J\ . \nonumber
\end{align}
The relation between the triple index $s$, $m$ and $n$, and the joined index $j$ is unique. Both index notations will be used throughout this paper, as the presented derivations benefit form either one or the other notation. The interested reader is referred to \cite{hansenSphericalNearfieldAntenna1988a,strattonElectromagneticTheory1941} for the actual physical meaning of $s$, $m$ and $n$. 

If reciprocity holds true for the antenna array under test (AUT), its transmission and reception coefficients are related by \cite{hansenSphericalNearfieldAntenna1988a}
\begin{align}
    R_{smn}=(-1)^m T_{s(-m)n}\ . \label{eq:TX_RX_coefs_reciprocity}
\end{align}
Provided that the AUT is connected to impedance-matched receivers, the outward propagating voltage waves $w^{(l)}$ at the antenna ports are
\begin{align}
    w^{(l)} = \sum\limits_{j=1}^{J} R_{j}^{(l)} a_j\ . \label{eq:single_antenna_RX}
\end{align} 
Relation \eqref{eq:single_antenna_RX} is a narrowband model. In the wideband case $w^{(l)}$, $R_{j}^{(l)}$ and $a_j$ change with frequency.

Our paper targets the optimization of antenna arrays for localizing sources in the far-field domain of the AUT. As such, the electromagnetic waves incident on the AUT are considered to be plane waves. We utilize the wideband plane wave model provided in Appendix \ref{sec:wideband_plane_wave_model}, with the respective incident SWCs 
\begin{align}
    a_{smn}(\omega, \theta_0, \varphi_0) = -A\frac{(-1)^m e^{-i\omega\tau}}{2r_0\sqrt{\pi Z_c}}  & S(\omega - \omega_0) \label{eq:wideband_asmn} \\
    & \myVec{P}^H \myVec{K}_{s(-m)n}(\theta_0, \varphi_0)\ . \nonumber
\end{align}
In \eqref{eq:wideband_asmn}, $r_0\in\mathbb{R}$, $\theta_0\in\mathbb{R}$ and $\varphi_0\in\mathbb{R}$ describe the position of the transmitter to be localized, in spherical coordinates relative to the AUT. Vector $\myVec{P}\in\mathbb{C}^3$ denotes the polarization of the incident wave as seen from the AUT, and $S(\omega)$ is the spectrum of the baseband signal at the transmitter side. Its equivalent time-domain signal, $s(t)=\mathcal{F}^{-1}\left\{ S(\omega) \right\}$ with $\mathcal{F}^{-1}$ denoting the inverse Fourier transform, is a real-valued low-pass signal at bandwidth $B$ given in $\si{\radian\per\second}$. Parameter $\omega_0$ is the angular carrier frequency of the RF channel in $\si{\radian\per\second}$. The propagation delay between source and AUT is denoted by $\tau=\frac{r_0}{c_0}$, factor $A\in\mathbb{R}$ models the signal amplitude, and $Z_c$ is the characteristic impedance of the transmission lines attached to the antenna ports.\footnote{We assume the same characteristic impedance at all antenna ports for simplicity, but all derivations generalize well for unequal characteristic impedances.} The \textit{Far-Field Functions} $\myVec{K}_{smn}$ represent the VSWFs $\myVec{F}_{smn}^{(3)}$ "far away" from the AUT:
\begin{align}
    \myVec{K}_{smn}(\theta, \varphi) = \lim_{kr\rightarrow\infty}\sqrt{4\pi}\frac{kr}{e^{ikr}} \myVec{F}_{smn}^{(3)} (kr, \theta, \varphi) \label{eq:far_field_pattern_functions}
\end{align}
The passband SWCs \eqref{eq:wideband_asmn} have non-zero components only within $-\omega_0 - B \leq \omega \leq -\omega_0 + B$, which may look unfamiliar at first hand. But as described in \cite{soklicFullSphereAntennaMeasurements2024}, \cite{IEEERecommendedPractice2012} and also Appendix \ref{sec:wideband_plane_wave_model}, this is a consequence of the VSWFs in \eqref{eq:VSWE_general} historically being defined with respect to a $e^{-i\omega t}$ dependency of the respective time-domain signals \cite{strattonElectromagneticTheory1941}. In other words, the VSWFs represent signals at negative frequencies. 

Inserting \eqref{eq:wideband_asmn} in \eqref{eq:single_antenna_RX} delivers the passband representations of the antenna port voltages
\begin{align}
    w^{(l)}(\omega) = -A\sum\limits_{smn} \bigg[ &R_{smn}^{(l)}(\omega) \frac{(-1)^m e^{-i\omega\tau}}{2r_0\sqrt{\pi Z_c}}  S(\omega - \omega_0) \label{eq:single_antenna_RX_with_SWCs:tmp} \nonumber \\
    &\myVec{P}^H \myVec{K}_{s(-m)n}(\theta_0, \varphi_0) \bigg]\ .
\end{align}
To obtain the equivalent baseband representation of \eqref{eq:single_antenna_RX_with_SWCs:tmp}, the signals are up-converted first with a carrier frequency of $\omega_0$, followed by a complex conjugation in the time-domain. If only an up-conversion is applied, the resulting baseband spectra would still be "mirrored" around $\omega=0$ compared to conventional baseband representations of passband signals. 
Applying the two operations delivers the final baseband representations of the antenna port voltages
\begin{flalign}
    &\tilde{w}^{(l)}(\omega) := \left\{w^{(l)}(-\omega-\omega_0)\right\}^* = \label{eq:single_antenna_RX_wideband} &\\
    &\quad \quad \quad\ = \sum\limits_{smn} \bigg[ \tilde{R}_{smn}^{(l)}(\omega) \frac{-A(-1)^m e^{-i(\omega+\omega_0)\tau}}{2r_0\sqrt{\pi Z_c}} \nonumber &\\
    &\quad \quad \quad \quad \quad \quad \quad \quad S(\omega) \myVec{P}^T \myVec{K}_{s(-m)n}^* (\theta_0, \varphi_0) \bigg]\nonumber \ , &
\end{flalign}
with $\tilde{R}_{smn}^{(l)}(\omega) := \left\{R_{smn}^{(l)}(-\omega - \omega_0)\right\}^*$ and considering that $S^*(-\omega) = S(\omega)$ due to $s(t)$ being a real-valued signal.

A wideband receiver attached to the port of antenna $l$ in the array samples the equivalent time-domain signals of \eqref{eq:single_antenna_RX_wideband} at samplingrate $f_s$. Assuming aliasing-free sampling with $f_s > \frac{B}{\pi}$, the obtained discrete frequency spectrum after applying the discrete Fourier transform to the sampled signals is
\begin{align}
    \tilde{w}^{(l)}[p] := \tilde{w}^{(l)}(p\Delta \omega)\ , \label{eq:continuous_to_discrete_frequency}
\end{align}
with $p \in \left[\frac{-P+1}{2}, \frac{P-1}{2}\right]$, $\Delta \omega = \frac{2\pi f_s}{P}$ and assuming that $P$ is an odd number.
The $P$ discrete frequency samples of all $L$ receivers attached to the array are gathered in a single vector $\myVec{\tilde{w}}\in \mathbb{C}^{PL}$. Utilizing the matrix and vector definitions from Appendix \ref{sec:matrix_vector_definitions}, $\myVec{\tilde{w}}$ can be written in matrix-vector notation as
\begin{align}
    \myVec{\tilde{w}} = \left(\myVec{1}_{L} \otimes \myVec{\tau}(\tau)\right) \odot \left(\myVec{R} \myVec{M}\myVec{K}^H(\theta_0, \varphi_0) \myVec{P}\right) \odot \left(\myVec{1}_{L} \otimes \myVec{S}\right)\ , \label{eq:RX_signal_vector_noise_free}
\end{align}
with $\otimes$ and $\odot$ denoting Kronecker and Hadamard products respectively. Vector $\myVec{\tau}\in\mathbb{C}^{P}$ contains the discrete frequency samples of phasor $e^{-i(\omega+\omega_0)\tau}$, matrix $\myVec{R}\in\mathbb{C}^{LP\times J}$ collects the array reception coefficients $R_{j}^{(l)}(p\Delta\omega)$ from \eqref{eq:single_antenna_RX_wideband}, matrix $\myVec{M}\in\mathbb{R}^{J\times J}$ is a diagonal matrix containing the $(-1)^m$ terms, and $\myVec{K}\in\mathbb{C}^{3\times J}$ is the matrix of far-field functions. Vector $\myVec{S}\in\mathbb{C}^{P}$ collects the discrete frequency samples of $S(\omega)$, multiplied with the remaining proportionality factors, and vector $\myVec{1}_{L}$ is a real-valued vector of length $L$ filled with ones.

The signal vector $\myVec{\tilde{w}}$ from \eqref{eq:RX_signal_vector_noise_free} depends on the three real-valued parameters $\tau$, $\theta_0$ and $\varphi_0$, denoting delay and DoA of the transmitter relative to the center of the AUTs circumscribing sphere. Additionally, $\myVec{\tilde{w}}$ depends on the polarization $\myVec{P}$, which is a complex-valued vector. In spherical coordinates, one can represent $\myVec{P}$ by means of four real-valued parameters $P_\theta$, $P_\varphi$, $\phi_\theta$ and $\phi_\varphi$ using 
\begin{align}
    \myVec{P} = P_\theta e^{i\phi_\theta} \myVec{i}_{\theta} + P_\varphi e^{i\phi_\varphi} \myVec{i}_{\varphi} \label{eq:polarization_vector}\ .
\end{align} 
Vectors $\myVec{i}_{\theta}, \myVec{i}_{\varphi} \in\mathbb{R}^{3}$ are unit vectors in elevation and azimuth directions. Due to the incident wave being a plane wave, the radial component of $\myVec{P}$ is zero. 

The signal amplitude $A$ and the path loss $\frac{1}{r_0}$ present in \eqref{eq:single_antenna_RX_wideband}
are dropped as explicit parameters of $\myVec{\tilde{w}}$, as in case of the AWGN channel conditions considered in this paper, they only affect the signal-to-noise ratio (SNR) at the receivers. During CRLB derivations presented in Section \ref{sec:FIM_and_CRLBs}, the SNR is however modelled via the covariance matrix of the noise vector affecting the received signals.
Thus, we finally express $\myVec{\tilde{w}}$ in terms of the seven real-valued parameters
\begin{align}
    \myVec{\theta} := \left[\tau, \theta_0, \varphi_0, P_\theta, P_\varphi, \phi_\theta, \phi_\varphi\right]^T\ , \label{eq:parameter_vector}
\end{align}
and denote the dependence of \eqref{eq:polarization_vector} on \eqref{eq:parameter_vector} from a notational perspective by
\begin{align}
    \myVec{\tilde{w}} \overset{\triangle}{=} \myVec{\tilde{w}} \left( \myVec{\theta} \right)\ .
\end{align}
To summarize the derivations so far, we derived a wideband frequency-domain model \eqref{eq:single_antenna_RX_wideband} for the signals received at the ports of an UWB antenna array under consideration of the frequency, polarization and direction-dependent reception characteristics of the AUT. After sampling, the vector discrete frequency samples for all antennas in the AUT can be expressed in matrix vector notation as shown in \eqref{eq:RX_signal_vector_noise_free}. This discrete frequency model forms the foundation for the CRLBs derived later in this paper.

We close this section by shortly addressing the topic of array manifold ambiguities \cite{manikasDifferentialGeometryArray2004}, besides not being the primary objective of this paper. As shown in Appendix \ref{sec:manifold_ambiguities}, a necessary condition for the AUT to not suffer from Type-I manifold ambiguities is, that the matrix of reception coefficients $\myVec{R}$ in \eqref{eq:RX_signal_vector_noise_free} has full rank.

\section{Obtaining SWCs from Simulation}
\label{sec:SWCs_from_simulation}

\begin{figure}[htbp]
    \centering
    \includegraphics[width=\columnwidth]{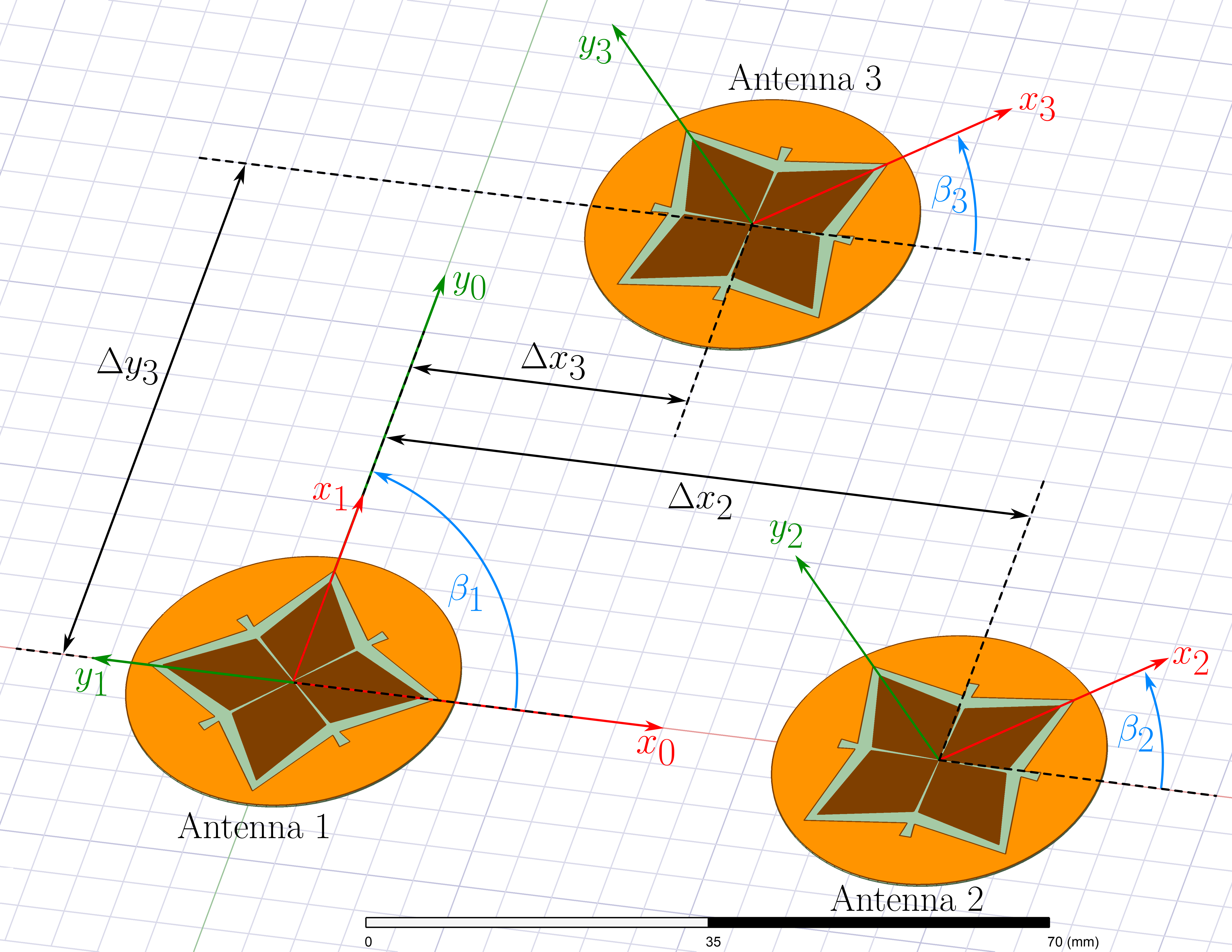}
    \caption{XETS antenna array to optimize. All three antennas can rotate in the $x_0y_0$-plane, with their rotation angles denoted by $\beta_1$, $\beta_2$ and $\beta_3$ respectively. Antenna 1 itself is only allowed to rotate. Antennas 2 and 3 are positioned relative to antenna 1, with their positions given by $\Delta x_2$, $\Delta x_3$ and $\Delta y_3$. Antenna 2 cannot move along the $y_0$ direction, to avoid collisions with antenna 3.}
    \label{fig:XETS_array_to_optimize}
\end{figure}

Evaluation of the signal model \eqref{eq:RX_signal_vector_noise_free} for a given AUT requires knowledge of the AUTs reception coefficients $R_{smn}^{(l)}$. This section describes a procedure for obtaining the $R_{smn}^{(l)}$ from the radiated field of a reciprocal AUT. Considering the example antenna array depicted Figure \ref{fig:XETS_array_to_optimize}, a known excitation voltage is applied to the port of one antenna in the array, while all other antenna ports are terminated with matched loads. For the currently active antenna, the electric field strength $\myVec{E}^{(l)}$ is exported on a spherical surface circumscribing the AUT. According to \cite{hansenSphericalNearfieldAntenna1988a}, the exported field strength $\myVec{E}^{(l)}$ can be expanded in terms of VSWFs as 
\begin{align}
    \myVec{E}^{(l)}(kr, \theta, \varphi) = \frac{k}{\sqrt{\eta}}\sum_{smn} v^{(l)} T_{smn}^{(l)} \myVec{F}_{smn}^{(3)}\left(kr, \theta, \varphi\right)\ , \label{eq:radiated_e_field}
\end{align}
as only radiated modes are present.
The array transmission coefficients $T_{smn}^{(l)}$ are obtained from the radiated field $\myVec{E}^{(l)}$ using
\begin{flalign}
    &\mathcal{N} := \int\limits_{\varphi=0}^{2\pi}\int\limits_{\theta=0}^{\pi}\left(\myVec{E}^{(l)}(kr, \theta, \varphi)\right)^T \myVec{F}_{smn}^{(3)}\left(kr, \theta, \varphi\right) \sin\theta d\theta d\varphi & \nonumber\\
    &\mathcal{D} := \frac{c_0 \sqrt{\eta}}{\omega} (-1)^{-m} \Bigg\{ & \nonumber\\
    &\quad \left(\mathcal{R}_{sn}^{(3)}(kr)\right)^2 +\delta_{s2} n(n+1)\left(\frac{\mathcal{R}_{2n}^{(3)}(kr)}{kr} \right)^2 \Bigg\} & \nonumber\\[1em]
    &T_{smn}^{(l)} = \frac{\mathcal{N}}{\mathcal{D}} \ .& \label{eq:array_TX_coefficients_from_simulation}
\end{flalign}
The integration over $\theta$ and $\varphi$ in the numerator $\mathcal{N}$ is carried out numerically, which requires $\myVec{E}^{(l)}$ to be sampled on a sufficiently dense grid along $\theta$ and $\varphi$.
The desired receive coefficients $R_{smn}^{(l)}$ are finally determined from the $T_{smn}^{(l)}$ via the reciprocity relation \eqref{eq:TX_RX_coefs_reciprocity}.  Relation \eqref{eq:array_TX_coefficients_from_simulation} exploits the orthogonality condition of VSWFs \cite{hansenSphericalNearfieldAntenna1988a}:
\begin{flalign}
    &\int\limits_{\varphi=0}^{\pi} \int\limits_{\theta=0}^{2\pi} \left(\myVec{F}^{(c)}_{smn}(kr, \theta,\varphi)\right)^T \myVec{F}^{(\gamma)}_{\sigma\mu\nu}(kr, \theta,\varphi) \sin\theta d\theta d\varphi = \nonumber &\\
    &\quad = \delta_{s\sigma,m(-\mu),n\nu} (-1)^m \Bigg\{ &\\
    &\quad \quad \mathcal{R}_{sn}^{(c)}(kr) \mathcal{R}_{sn}^{(\gamma)}(kr) +\delta_{s2} n(n+1)\frac{\mathcal{R}_{2n}^{(c)}(kr)}{kr} \frac{\mathcal{R}_{2n}^{(\gamma)}(kr)}{kr} \Bigg\}\nonumber\ , \label{eq:VSWF_orthogonality}&
\end{flalign}
where $\mathcal{R}_{sn}^{(c)}$ are the spherical Bessel functions describing the radial dependency of the VSWFs, and
\begin{align}
    \delta_{s\sigma,m(-\mu),n\nu} = \begin{cases}
        1 & s=\sigma, m=-\mu, n=\nu \\
        0 & \text{else}\ .
    \end{cases}
\end{align}

The attentive reader may have noticed that the issue of the $e^{-i\omega t}$ timebasis of the VSWFs has not been addressed so far, although this issue received quite some attention in Section \ref{sec:wideband_signal_model}. 
That is because the majority of EM simulators deliver field results for $\omega > 0$ only, as their spectra are Hermitian symmetric because the respective time-domain quantities are real-valued. The Hermitian symmetry allows to directly expand the radiated field $\myVec{E}^{(l)}$ by means of \eqref{eq:radiated_e_field}, without requiring to correct for the $e^{-i\omega t}$ timebasis of the VSWFs, as
\begin{gather*}
    \myVec{e}^{(l)}(t) := \mathcal{F}^{-1}\left\{ \myVec{E}^{(l)}(\omega) \right\},\ \ \myVec{\hat{e}}^{(l)}(t) := \left[ \myVec{e}^{(l)}(t) \right]^* \Rightarrow \\[0.25em]
    \myVec{\hat{E}}^{(l)}(\omega) := \mathcal{F}\left\{ \myVec{\hat{e}}^{(l)}(t) \right\} = \left(\myVec{E}^{(l)}(-\omega)\right)^* = \myVec{E}^{(l)}(\omega)\ .
\end{gather*}

\section{Fisher Information Matrix and Cram\'er-Rao Lower Bounds}
\label{sec:FIM_and_CRLBs}

In a real-world scenario, the signals obtained from the $L$ receive antennas are affected by noise. 
Considering additive Gaussian noise $\myVec{n}$ with covariance matrix $\myVec{C}_n$, the resulting discrete-frequency signal vector is
\begin{align}
    \myVec{x}(\myVec{\theta}) := \myVec{\tilde{w}}(\myVec{\theta}) + \myVec{n}\ . \label{eq:RX_signal_vector_with_noise}
\end{align}
The likelihood of $\myVec{x}$ with respect to parameters $\myVec{\theta}$ is  
\begin{align}
    p_{\myVec{x}}\left(\myVec{x};\myVec{\theta}\right) = \frac{1}{\pi^{LP}\left|\myVec{C}_{\myVec{n}}\right|} \exp\left\{ -\left(\myVec{x} - \myVec{\tilde{w}}(\myVec{\theta}) \right)^H \myVec{C}_{\myVec{n}}^{-1} \left(\myVec{x} - \myVec{\tilde{w}}(\myVec{\theta}) \right) \right\}\ .\label{eq:AWGN_signal_vector_likelihood}
\end{align}
For likelihoods of the form \eqref{eq:AWGN_signal_vector_likelihood}, the FIM $\myVec{F}$ is obtained, according to \cite{kayFundamentalsStatisticalSignal1993a}, from
\begin{align}
    \left[\myVec{F}\left(\myVec{\theta}\right)\right]_{kl} = 2\text{Re}\left\{ \frac{\partial \myVec{\tilde{w}}^H(\myVec{\theta})}{\partial \theta_k} \myVec{C}_{\myVec{n}}^{-1}  \frac{\partial \myVec{\tilde{w}}(\myVec{\theta})}{\partial \theta_l} \right\}\ . \label{eq:FIM}
\end{align}
The crucial part of \eqref{eq:FIM} is the calculation of the partial derivatives of  $\myVec{\tilde{w}}$ with respect to the signal parameters $\myVec{\theta}$. Their derivations are presented in detail in Appendix \ref{sec:FIM_partial_derivatives}.
Once the FIM has been determined, the Cram\'er-Rao Lower Bounds bound $\mathcal{B}_{\theta_k}(\myVec{\theta})$ of a signal parameter $\theta_k$ is determined as
\begin{align}
    \text{var}\left\{\theta_k \right\}\left(\myVec{\theta}\right) \geq \mathcal{B}_{\theta_k}(\myVec{\theta}) := \left[\myVec{F}^{-1}\left(\myVec{\theta}\right)\right]_{kk} \ . \label{eq:CRLBs}
\end{align}

\section{Formulating an Optimization Problem}
\label{sec:optimization_problems}
Upon changing positions and orientations of the antenna elements in an array, the array reception coefficients $\myVec{R}$ in \eqref{eq:RX_signal_vector_noise_free} will change as well, together with the FIM $\myVec{F}$ \eqref{eq:FIM} and the CRLBs $\mathcal{B}_{\theta_k}$ \eqref{eq:CRLBs}. As the CRLBs determine lower bounds for the estimation error variances of the respective signal parameters, an optimization problem can be formulated to determine the antenna positions and orientations that result in the best localization performance. For this purpose, the antenna positions and orientations are collected in a vector $\myVec{\gamma}\in\mathbb{R}^{N_\gamma}$, with $N_\gamma$ being the number of antenna positions and orientations considered as optimization parameters. We formally express the dependence of $\myVec{R}$, $\myVec{\tilde{w}}$, $\myVec{F}$ and $\mathcal{B}_{\theta_k}$ on $\myVec{\gamma}$ by
\begin{align*}
    \myVec{R} \rightarrow \myVec{R}(\myVec{\gamma})&,\ \myVec{\tilde{w}}(\myVec{\theta}) \rightarrow \myVec{\tilde{w}}\left(\myVec{\theta}, \myVec{\gamma}\right), \\[0.5em]
    \myVec{F}\left(\myVec{\theta}\right) \rightarrow \myVec{F}\left(\myVec{\theta}, \myVec{\gamma}\right)&,\  \mathcal{B}_{\theta_k}\left(\myVec{\theta}\right) \rightarrow \mathcal{B}_{\theta_k}\left(\myVec{\theta}, \myVec{\gamma}\right)\ .
\end{align*} 

An obvious optimization criterion is now to minimize the sum of CRLBs $\mathcal{B}_{\theta_k}$, or in other words, the trace of the inverse FIM:
\begin{align}
    J_A(\myVec{\gamma}) = \int\limits_{\Omega_{\myVec{\theta}}} \sum\limits_{k=1}^{7} \mathcal{B}_{\theta_k}\left(\myVec{\theta}, \myVec{\gamma}\right) d\myVec{\theta} = \int\limits_{\Omega_{\myVec{\theta}}} \text{tr}\left\{\myVec{F}^{-1}\left(\myVec{\theta}, \myVec{\gamma}\right)\right\} d\myVec{\theta} \label{eq:A_optimality_criterion}
\end{align}
The volume $\Omega_{\myVec{\theta}} \subset \mathbb{R}^7$ is the domain of all possible values for the signal parameters $\myVec{\theta}$. It is given by
\begin{align*}
    \tau\in\left[0, \tau_{\text{max}}\right] &,\ \theta \in \left[0, \pi \right],\ \\
    \varphi, \phi_\theta, \phi_\varphi \in\left[0, 2\pi\right]&,\ P_\theta,P_\varphi \in\left[0,1 \right]\ ,
\end{align*} 
subject to the constraint that polarization $\myVec{P}$ must be a unit vector, that is
\begin{align*}
    \sqrt{P_\theta^2 + P_\varphi^2}\overset{!}{=}1\ .
\end{align*}
The maximum propagation delay $\tau_{\text{max}}$ is determined by the sample spacing $\Delta \omega$ introduced in \eqref{eq:continuous_to_discrete_frequency}. An unambiguous delay estimation is only possible if the phase between two consecutive frequency domain samples does not change by more than $2\pi$, that is
\begin{align}
    \Delta \omega \tau < 2\pi \Rightarrow \tau < \tau_{\text{max}}=\frac{2\pi}{\Delta \omega}\ . \label{eq:unambiguous_delay_condition}
\end{align}
Based on the objective function \eqref{eq:A_optimality_criterion} an optimization problem is now defined as:
\begin{gather}
    \myVec{\hat{\gamma}}_A = \underset{\myVec{\gamma}}{\text{min}}\ J_A(\myVec{\gamma})  \label{eq:A_optimization} \\
    \text{s.t.}\ \myVec{g}(\myVec{\gamma}) \leq 0\ \& \ \myVec{h}(\myVec{\gamma})=\myVec{0} \ . \nonumber  
\end{gather}
The $N_h$ equality constraints $\myVec{h}(\myVec{\gamma})\in \mathbb{R}^{N_h}$, together with the $N_g$ inequality constraints $\myVec{g}(\myVec{\gamma})\in \mathbb{R}^{N_g}$, restrict the possible positions and orientations of the antenna elements. In the majority of cases, the antenna positions and orientations $\myVec{\gamma}$ are subject to simple boundary constraints, which are
\begin{align*}
    \myVec{\gamma}_{l} \leq \myVec{\gamma} \leq \myVec{\gamma}_u\ ,
\end{align*}
with $\myVec{\gamma}_{l}$ and $\myVec{\gamma}_u$ being the respective lower and upper antenna parameter bounds.

An optimization problem of the form \eqref{eq:A_optimization} minimizes the average variance of the parameter estimates $\hat{\theta}_k$, commonly known as \textit{A-Optimality Criterion} in the design of statistical experiments \cite{atkinsonOptimumExperimentalDesigns2007,pukelsheimOptimalDesignExperiments2006a}. A very popular alternative to the A-optimality criterion is the \textit{D-Optimality Criterion}, where the determinant of the inverse FIM is minimized instead of its trace. This minimizes the generalized variance instead of the average variance of the parameter estimates. The D-optimality criterion also allows to define an objective function without requiring to invert the FIM:
$\left|\myVec{F}^{-1}\right| \equiv \left|\myVec{F}\right|^{-1}$:
\begin{gather}
    J_D(\myVec{\gamma}) = -\int\limits_{\Omega_{\myVec{\theta}}} \left|\myVec{F}\left(\myVec{\theta}, \myVec{\gamma}\right)\right|d\myVec{\theta}\ , \label{eq:D_optimality_criterion}
\end{gather} 
The resulting optimization problem is 
\begin{gather}
    \myVec{\hat{\gamma}}_D = \underset{\myVec{\gamma}}{\text{min}}\ J_D(\myVec{\gamma})  \label{eq:D_optimization} \\
    \text{s.t.}\ \myVec{g}(\myVec{\gamma}) \leq 0\ \& \ \myVec{h}(\myVec{\gamma})=\myVec{0} \ . \nonumber  
\end{gather}
The D-optimality criterion results in a lower computational effort for the calculation of the objective function, as the FIM inversion required in \eqref{eq:A_optimality_criterion} is avoided.
Also, possible issues with singular FIMs for certain $\myVec{\theta}$ or $\myVec{\gamma}$ are avoided.
\section{Example: Optimal 3-element XETS Array for Linearly Polarized Incident Waves}
\label{sec:example}
The proposed optimization procedure is demonstrated by means of the planar three-element array depicted in Figure \ref{fig:XETS_array_to_optimize}. The array elements are \textit{Crossed Exponentially Tapered Slot (XETS)} antennas introduced in \cite{costaPerformanceCrossedExponentially2009}.
We aim to find the positions and orientations of antennas 2 and 3 relative to antenna 1, in order to optimize the achievable localization performance in terms of the D-optimality criterion \eqref{eq:D_optimization}. The antennas are only allowed to rotate in the $xy$-plane, which results in a total of 6 optimization parameters
\begin{align}
    \myVec{\gamma} = \left[\Delta x_2, \Delta x_3, \Delta y_3, \beta_1, \beta_2, \beta_3 \right]\ . \label{eq:example_problem_parameter_vector}
\end{align}
The search space for $\myVec{\gamma}$ is determined by
\begin{align}
    \Delta y_3, \Delta x_2 &\in \left[35\si{\milli\meter}, 70\si{\milli\meter}\right] \label{eq:example_boundary_constraints}\\ 
    \Delta x_3 &\in \left[0\si{\milli\meter}, 70\si{\milli\meter}\right] \\ \nonumber
    \beta_1, \beta_2, \beta_3 &\in \left[0\si{\degree}, 180\si{\degree} \right]\ , \nonumber
\end{align}
The lower limits of $\Delta x_2$, $\Delta x_3$ and $\Delta y_3$ are selected such that the antenna elements cannot overlap.
Due to the rotational symmetry of the XETS antenna elements, only the $\left[0\si{\degree}, 180\si{\degree} \right]$ interval needs to be considered for the antenna orientations $\beta_1$, $\beta_2$ and $\beta_3$.

For demonstration purposes, only linearly polarized waves are considered to be incident on the AUT. This restriction reduces the dimension of the FIM $\myVec{F}$ from \eqref{eq:FIM} from a $7\times 7$ to a $4 \times 4$ matrix, as the four polarization parameters $P_\theta$, $P_\varphi$, $\phi_\theta$ and $\phi_\varphi$ reduce to a single parameter $\alpha \in \left[0, \pi\right]$, expressing the slant of the polarization plane:
\begin{align*}
    \myVec{P} &= P_\theta e^{i\phi_\theta} \myVec{i}_{\theta} + P_\varphi e^{i\phi_\varphi} \myVec{i}_{\varphi}
    \rightarrow \\
    \myVec{P} &=  \sin(\alpha) \myVec{i}_{\theta} + \cos(\alpha) \myVec{i}_{\varphi}\ .
\end{align*}
One obtains $\phi_\theta = \phi_\varphi \equiv 0$, $P_\theta = \sin(\alpha)$ and $P_\varphi=\cos(\alpha)$\ .
The modified signal parameter vector $\myVec{\eta} \in \mathbb{R}^4$ is obtained as
\begin{align*}
    \myVec{\eta} = \left[\tau, \theta_0, \varphi_0, \alpha\right]^T \ .
\end{align*}
Modifying the parameter vector also requires to re-parameterize the FIM $\myVec{F}$ from \eqref{eq:FIM} with respect to $\myVec{\eta}$. For this purpose the Jacobian of the original parameter vector $\myVec{\theta}$ from \eqref{eq:parameter_vector} with respect to the new parameter vector $\myVec{\eta}$ is required \cite{lehmannTheoryPointEstimation1998}:
\begin{align*}
    \left[ \myVec{\mathcal{I}}_{\myVec{\eta}}\right]_{kl} = \frac{\partial \theta_k}{\partial \eta_{l}}\ .
\end{align*}
Fortunately, the calculations are straight forward, and one obtains
\begin{align*}
    \myVec{\mathcal{I}}_{\myVec{\eta}} = 
    \begin{bNiceArray}{c|c}[margin]
        \myVec{I}_{3 \times 3} & \myVec{0}_{3 \times 1} \\
        \hline \Block{4-1}{\myVec{0}_{4 \times 3}} & -\sin(\alpha) \\
                               & \cos(\alpha) \\
                               & 0 \\
                               & 0
    \end{bNiceArray}\ ,
\end{align*}
with $\myVec{I}_{N \times N}$ being an identity matrix of dimension $N\times N$, and $\myVec{0}_{N \times M}$ being a matrix of dimension $N\times M$ filled with zeros.
The re-parameterized FIM with respect to $\myVec{\eta}$ is then according to \cite{lehmannTheoryPointEstimation1998}
\begin{align}
    \myVec{F}_{\myVec{\eta}}(\myVec{\eta}) := \myVec{\mathcal{I}}_{\myVec{\eta}}^T (\myVec{\eta}) \myVec{F}\left(\myVec{\theta}(\myVec{\eta})\right) \myVec{\mathcal{I}}_{\myVec{\eta}}(\myVec{\eta})\ . \label{eq:FIM_linear_polarization}
\end{align}
The objective functions $J_A(\myVec{\gamma})$ and $J_D(\myVec{\gamma})$ from \eqref{eq:A_optimality_criterion} and \eqref{eq:D_optimality_criterion} for linearly polarized incident waves are obtained as
\begin{align}
    J_{A, \text{lin}}(\myVec{\gamma}) &= \int\limits_{\Omega_{\myVec{\eta}}} \text{tr}\left\{\myVec{F}_{\myVec{\eta}}^{-1}\left(\myVec{\eta}, \myVec{\gamma}\right)\right\} d\myVec{\eta} \label{eq:objective_functions_linear_polarization}\\
    J_{D, \text{lin}}(\myVec{\gamma}) &= -\int\limits_{\Omega_{\myVec{\eta}}} \left|\myVec{F}_{\myVec{\eta}}(\myVec{\eta}, \myVec{\gamma})\right|d\myVec{\eta}\ , \nonumber
\end{align}
with $\Omega_{\myVec{\eta}}$ given by 
\begin{align*}
    \tau\in\left[0, \tau_{\text{max}}\right] ,\ \theta, \alpha \in \left[0, \pi \right],\ \varphi\in\left[0, 2\pi\right]\ .
\end{align*} 

The AUT from Figure \ref{fig:XETS_array_to_optimize} is optimized for an operation at Channel 9 of the \textit{HRP UWB} PHY-layer defined in \cite{9144691}, which offers $\approx 500\si{\mega\hertz}$ bandwidth at a center frequency of $\approx 8\si{\giga\hertz}$. A full-wave EM simulation is carried out at $21$ equidistant frequency points within the interval $\left[7.75\si{\giga\hertz}, 8.25\si{\giga\hertz}\right]$. For each frequency bin, the array reception coefficients $R_{smn}^{(l)}$ are determined from the radiated fields of the three antenna elements as described in Section \ref{sec:SWCs_from_simulation}. The VSWE was thereby truncated at a maximum order of $N=25$, leading to $J=1350$ modes considered per frequency sample per antenna element. The total number of modes considered for each antenna array during the optimization is thus $3\cdot21\cdot1350 =85050$.
The reception coefficients are then applied in \eqref{eq:RX_signal_vector_noise_free} to obtain the signal vector $\tilde{\myVec{w}}$. For demonstration purposes, vector $\myVec{S}$ in \eqref{eq:RX_signal_vector_noise_free}, representing the spectrum of the baseband signal at the transmitter, is considered to be of the form $\myVec{S} = \myVec{1}_{P}$. This corresponds to a sinc-pulse in time-domain, with proper normalization such that the magnitude scaling factors in \eqref{eq:matrix_vector_definitions} vanish. 
Last but not least, the noise covariance matrix $\myVec{C}_n$ required in \eqref{eq:FIM} is considered to be of the form $\myVec{C}_n = \sigma_n^2 \myVec{I}_{LP\times LP}$, with $\sigma_n^2=0.01$. This has the advantageous side effect that only one particular value for the propagation delay $\tau$ needs to be considered during the optimization, as shown below. This shrinks the integration domain for the objective functions in \eqref{eq:objective_functions_linear_polarization} and thus further reduces computational effort for calculating the objective functions. For the selected $\myVec{C}_n$, expression \eqref{eq:FIM} for calculating the FIM $\myVec{F}$ can be rephrased to
\begin{align*}
    \left[\myVec{F}\left(\myVec{\theta}\right)\right]_{kl} = \frac{2}{\sigma_n^2}\text{Re}\left\{ \frac{\partial \myVec{\tilde{w}}^H(\myVec{\theta})}{\partial \theta_k}\frac{\partial \myVec{\tilde{w}}(\myVec{\theta})}{\partial \theta_l} \right\}\ .
\end{align*}
Considering the partial derivatives of $\myVec{\tilde{w}}$ from Appendix \ref{sec:FIM_partial_derivatives}, the FIM entries are of the form
\begin{align}
    \left[\myVec{F}\left(\myVec{\theta}\right)\right]_{kl} = \|\myVec{1}_{L\times 1} \otimes \myVec{\tau}\|^2 \odot \|\myVec{1}_{L\times 1} \otimes \myVec{S}\|^2 \odot \mathcal{T}\left(\theta_0, \varphi_0, \myVec{P} \right) \ \label{eq:FIM_uncorrelated_noise},
\end{align}
with $\mathcal{T}$ being a real-valued term denoting the dependence of the FIM with respect to DoA and polarization.
The propagation delay $\tau$ enters \eqref{eq:FIM_uncorrelated_noise} only in vector $\myVec{\tau}$ in the first term. Vector $\myVec{\tau}$ itself is just a vector of complex exponentials according to \eqref{eq:matrix_vector_definitions}, so the term $\|\myVec{1}_{L\times 1} \otimes \myVec{\tau}\|^2$ has a constant value independent of $\tau$. The FIM $\myVec{F}$ is therefore constant for arbitrary values of $\tau$. This property is also not affected through the re-parameterization of the FIM in \eqref{eq:FIM_linear_polarization}.
A second consequence of our selection of $\myVec{C}_n$ is, that the spectrum of the TX pulse shape $S(\omega)$ enters via $\| \myVec{1}_{L\times 1} \otimes \myVec{S} \|^2$. As such, different TX pulse shapes only result in different scaling factors applied to the entire FIM.

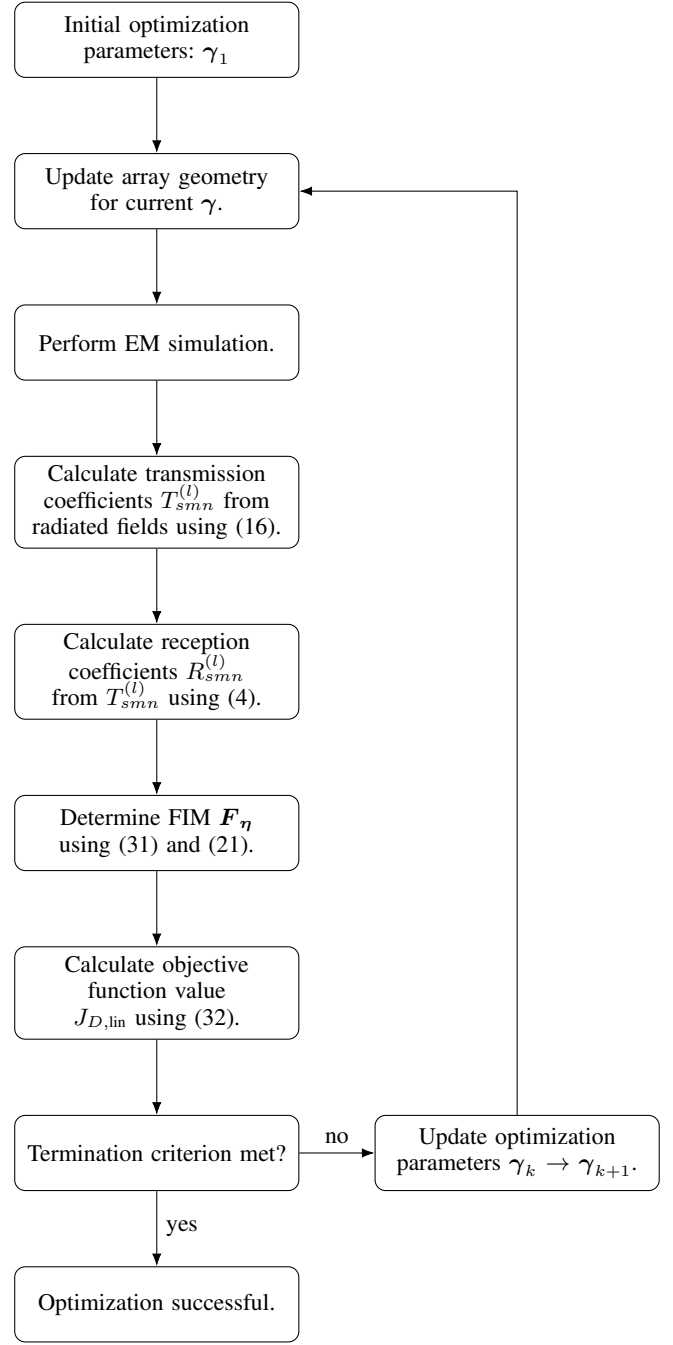
\begin{figure}
    \centering
    \begin{tikzpicture}[
        block/.style={
        draw, 
        fill=white, 
        text width=0.4*\columnwidth, 
        minimum height=1cm,
        rounded corners,
        align=center
        }, 
        font=\small
        ]
        \node[block] (start) {Initial optimization parameters: $\myVec{\gamma}_1$};
        \node[block] (step1) [below=of start] {Update array geometry for current $\myVec{\gamma}$.};
        \node[block] (step2) [below=of step1] {Perform EM simulation.};
        \node[block] (step3) [below=of step2] {Calculate transmission coefficients $T_{smn}^{(l)}$ from radiated fields using \eqref{eq:array_TX_coefficients_from_simulation}.};
        \node[block] (step4) [below=of step3] {Calculate reception coefficients $R_{smn}^{(l)}$ from $T_{smn}^{(l)}$ using \eqref{eq:TX_RX_coefs_reciprocity}.};
        \node[block] (step5) [below=of step4] {Determine FIM $\myVec{F}_{\myVec{\eta}}$ using \eqref{eq:FIM_linear_polarization} and \eqref{eq:FIM}.};
        \node[block] (step6) [below=of step5] {Calculate objective function value $J_{D, \text{lin}}$ using \eqref{eq:objective_functions_linear_polarization}.};
        \node[block] (step7) [below=of step6] {Termination criterion met?};
        \node[block] (step8) [right=of step7] {Update optimization parameters $\myVec{\gamma}_k \rightarrow \myVec{\gamma}_{k+1}$.};
        \node[block] (step9) [below=of step7] {Optimization successful.};

        \node[above] at ($(step7.east)!0.5!(step8.west)$) {no};
        \node[right] at ($(step7.south)!0.5!(step9.north)$) {yes};

        \draw[-Latex] (start.south) -- (step1.north);
        \draw[-Latex] (step1.south) -- (step2.north);
        \draw[-Latex] (step2.south) -- (step3.north);
        \draw[-Latex] (step3.south) -- (step4.north);
        \draw[-Latex] (step4.south) -- (step5.north);
        \draw[-Latex] (step5.south) -- (step6.north);
        \draw[-Latex] (step6.south) -- (step7.north);
        \draw[-Latex] (step7.east) -- (step8.west);
        \draw[-Latex] (step7.south) -- (step9.north);
        \draw[-Latex] (step8.north) |- (step1.east);
    \end{tikzpicture}
    \caption{Flow-chart of the optimization procedure for the example 3-element XETS array.}
    \label{fig:optimization_flow_chart}
\end{figure}

The 3-element XETS array is optimized with respect to $J_{D, \text{lin}}$ objective function from \eqref{eq:objective_functions_linear_polarization}. The antenna positions and orientations from \eqref{eq:example_problem_parameter_vector} are thereby subject to the boundary constraints \eqref{eq:example_boundary_constraints}
\begin{align}
    \myVec{\gamma}_{\text{lin},l} &:= 
    \begin{bmatrix}
        35\si{\milli\meter} &
        0\si{\milli\meter} &
        35\si{\milli\meter} &
        0\si{\degree} &
        0\si{\degree} &
        0\si{\degree}
    \end{bmatrix}^T \label{eq:boundary_constraints}\\
    \myVec{\gamma}_{\text{lin},u} &:= 
    \begin{bmatrix}
        70\si{\milli\meter} &
        70\si{\milli\meter} &
        70\si{\milli\meter} &
        180\si{\degree} &
        180\si{\degree} &
        180\si{\degree}
    \end{bmatrix}^T\ ,\quad \nonumber
\end{align}
which delivers the final optimization problem formulation
\begin{gather}
    \myVec{\hat{\gamma}}_{\text{lin}} = \underset{\myVec{\gamma}}{\text{min}}\ J_{D, \text{lin}}(\myVec{\gamma})  \label{eq:D_optimization_example} \\
    \text{s.t.}\ \myVec{\gamma}_{\text{lin},l} \leq \myVec{\gamma} \leq \myVec{\gamma}_{\text{lin},l}\nonumber\ .
\end{gather}
The optimization problem \eqref{eq:D_optimization_example} is solved using the \textit{Differential Evolution} (DE) global optimization algorithm \cite{stornDifferentialEvolutionSimple1997}, as there is no guarantee that \eqref{eq:D_optimization_example} is a convex optimization problem. More specifically, the DE implementation provided by SciPy \cite{virtanenSciPy10Fundamental2020} has been used. A flow-chart summarizing the entire the optimization procedure is depicted in Figure \ref{fig:optimization_flow_chart}. 
It shall be noticed at this point that the position boundaries \eqref{eq:boundary_constraints} may result in an aliased antenna array being delivered by the optimization, considering that the freespace wavelength at UWB channel 9 is $\approx 38\si{\milli\meter}$. One could even expect an aliased array resulting from the optimization, as first, the CRLBs themselves do not account for possible manifold ambiguities and second, the AoA CRLBs for ideal antenna arrays decrease with increasing antenna spacing \cite{wildingAccuracyBoundsArrayBased2018}.  

\section{Results \& Discussion}
\label{sec:example_results}
\begin{figure}[htbp]
    \centering
    \begin{minipage}[b]{0.85\columnwidth}
        \centering
        \includegraphics[width=\columnwidth]{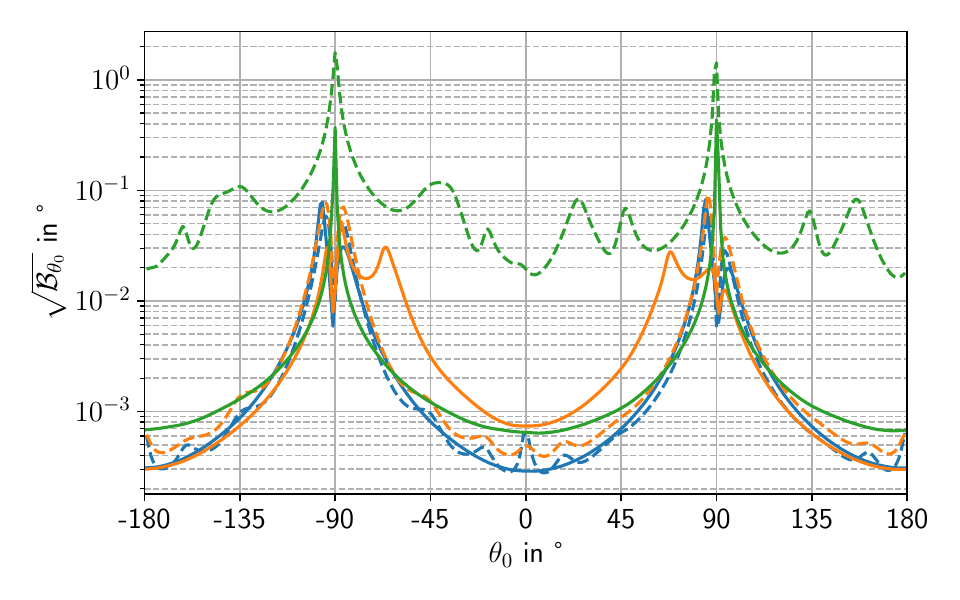}
        (a)
    \end{minipage}\\
    \begin{minipage}[b]{0.85\columnwidth}
        \centering
        \includegraphics[width=\columnwidth]{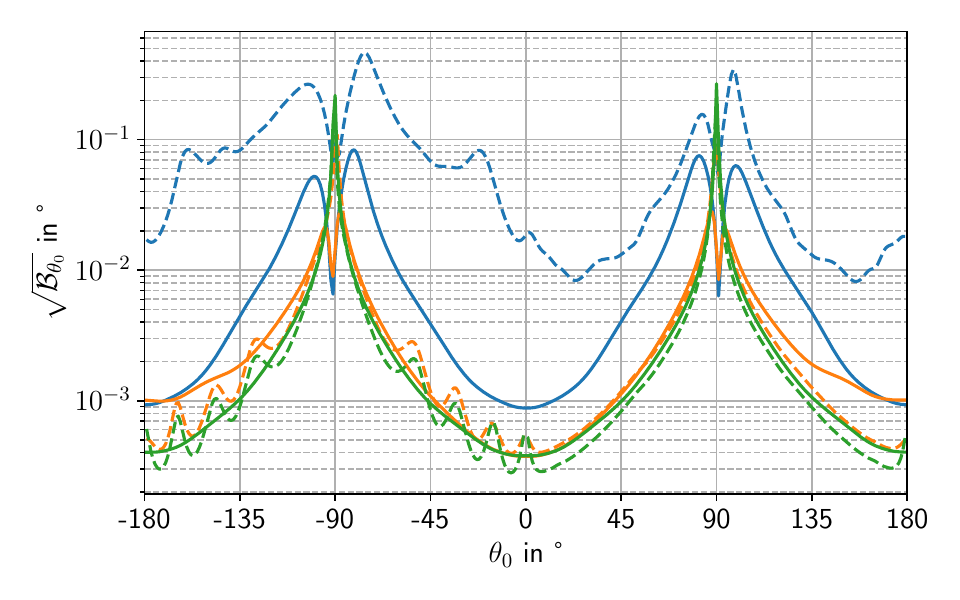}
        (b)
    \end{minipage}\\
    \begin{minipage}[b]{0.85\columnwidth}
        \centering
        \includegraphics[width=\columnwidth]{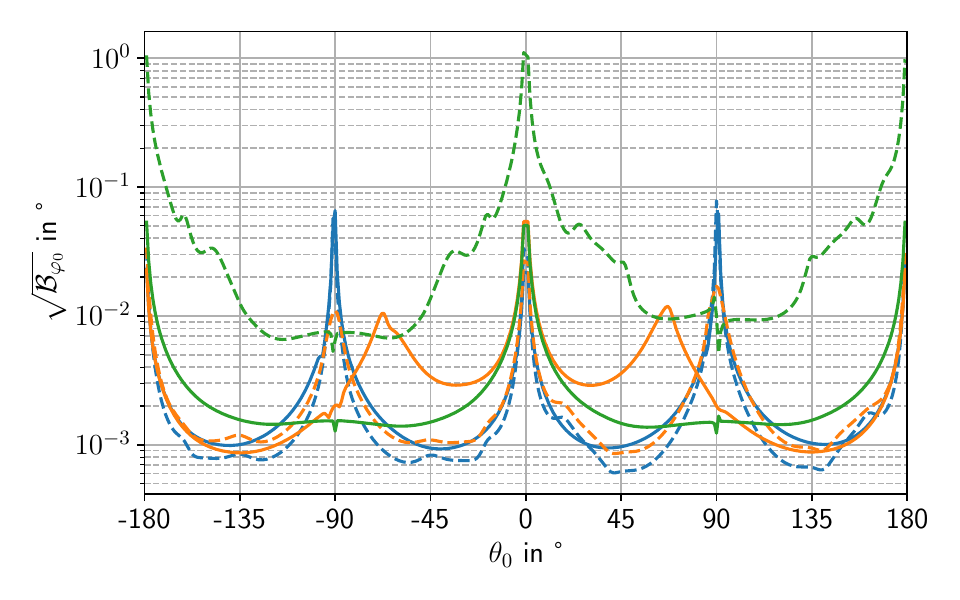}
        (c)
    \end{minipage}\\
    \begin{minipage}[b]{0.85\columnwidth}
        \centering
        \includegraphics[width=\columnwidth]{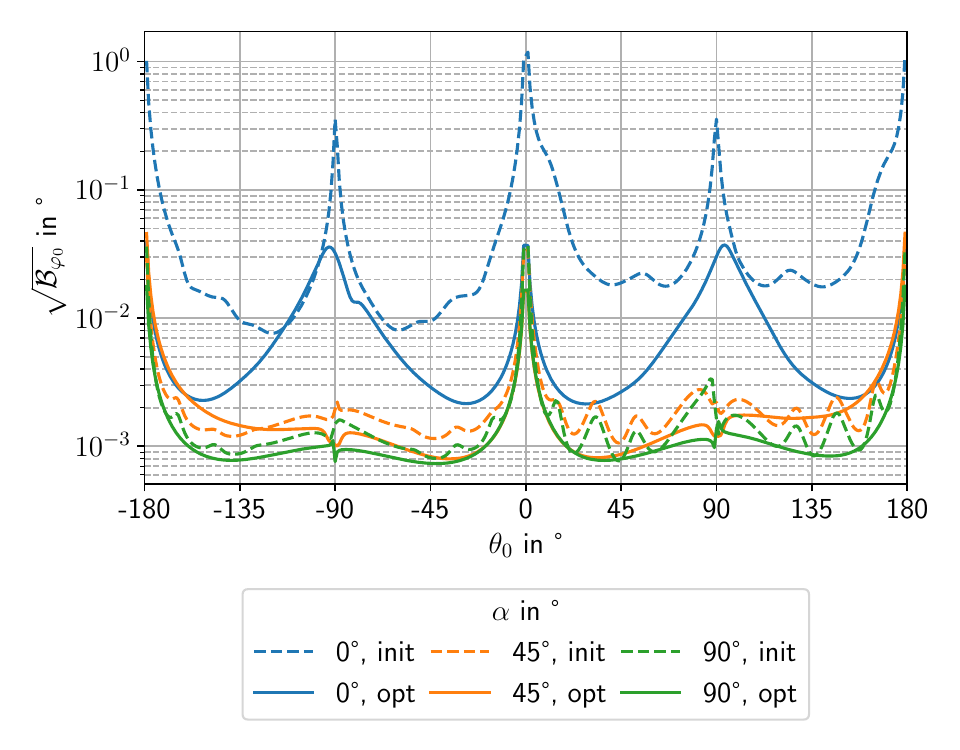}
        (d)
    \end{minipage}
    \caption{Comparison of the azimuth and elevation CRLBs, $\mathcal{B}_{\varphi_0}$ and $\mathcal{B}_{\theta_0}$, delivered by the initial and optimal antenna arrays. Figures (a) and (b) depict $\mathcal{B}_{\varphi_0}$, whereas (c) and (d) depict $\mathcal{B}_{\varphi_0}$. Dashed lines are the results for the initial array, solid lines are the results for the optimized array. Results for the $\varphi_0=0\si{\degree}$ principal cut, are shown in (a) and (c), whereas (b) and (d) depict the results for the$\varphi_0=90\si{\degree}$ principal cut.}
    \label{fig:CRLBs_initial_vs_optimized} 
\end{figure}

\begin{figure}[htbp]
    \centering
    \begin{minipage}[b]{0.75\columnwidth}
        \centering
        \includegraphics[width=\columnwidth]{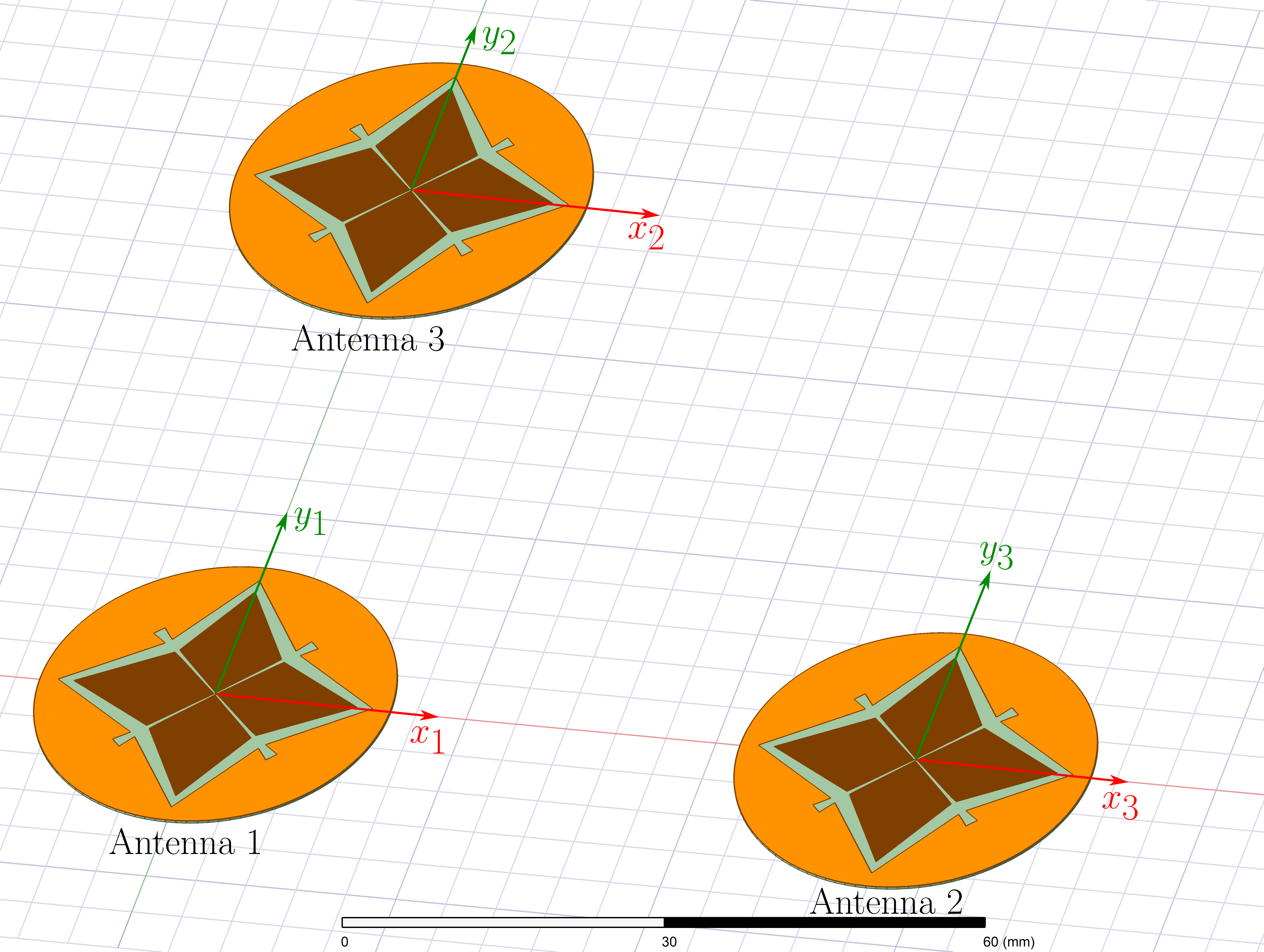}
        (a)
    \end{minipage}\\
    \begin{minipage}[b]{0.7\columnwidth}
        \centering
        \includegraphics[width=\columnwidth]{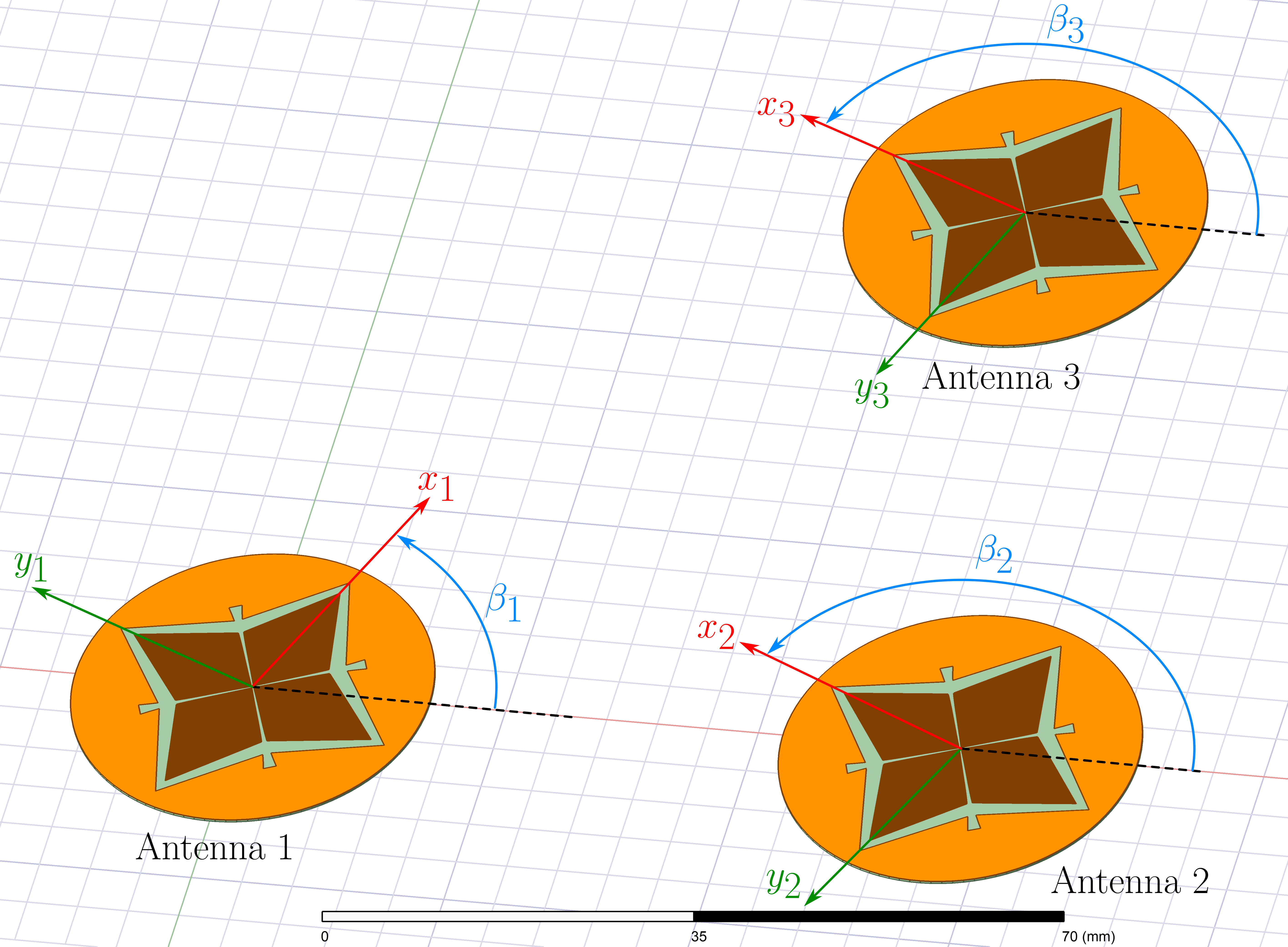} 
        (b)
    \end{minipage}
    \caption{(a) Initial antenna array versus (b) antenna array after optimization.}
    \label{fig:antenna_arrangements} 
\end{figure}

In each iteration of the optimization procedure, a population size of $18$ antenna arrays has been analyzed. The optimization procedure was executed for $40$ generations, leading to a total number $756$ different antenna arrays being analyzed. Ansys HFSS version 2025R1 was used as EM simulator. Simulating 5 antenna arrays in parallel and utilizing computational resources of 4 CPU cores and 8GB for each antenna array, the optimization procedure took four days to complete.
Figure \ref{fig:antenna_arrangements} depicts the initial antenna array considered for the optimization, together with the final antenna array after the optimization procedure has finished. The exact antenna positions and orientations for the initial and the optimal antenna arrays are found in Table \ref{tab:antenna_positions}. 
According to Table \ref{tab:antenna_positions}, the antenna element spacings of the optimized arrangement stayed similar or even increased compared to the initial arrangement. This matches the expectations from CRLB formulas for ideal antenna arrays \cite{wildingAccuracyBoundsArrayBased2018}, which show that AoA CRLBs become lower for larger antenna spacings. As such, one could expect that the antenna spacings of the XETS array would increase even further, if the upper limits for the antenna positions would be raised form the current value of $70\si{\milli\meter}$.\\
We further compare the initial and optimal antenna arrays by means of the CRLBs for the azimuth and elevation estimates of the incident wave direction, $\mathcal{B}_{\varphi_0}$ and $\mathcal{B}_{\theta_0}$, according to \eqref{eq:CRLBs}.
Figures \ref{fig:CRLBs_initial_vs_optimized}a to \ref{fig:CRLBs_initial_vs_optimized}d depict the two CRLBs over the true elevation $\theta_0$ for the two principal azimuth cuts at $\varphi_0=0\si{\degree}$ and $\varphi_0=90\si{\degree}$ \footnote{The VSWFs are defined for $\theta_0 \in \left[0\si{\degree}, 180\si{\degree}\right]$ in the coordinate system from \cite[Sec. 7]{9714428}. Quantities plotted for $\theta_0 \in \left[-180\si{\degree}, 0\si{\degree}\right]$ are therefore obtained from adding a $180\si{\degree}$ offset to the $\varphi_0$ value of the desired principal cut.}, and for three different polarization angles $\alpha\in\left\{0\si{\degree}, 45\si{\degree}, 90\si{\degree} \right\}$. 
To compare both antenna arrays also in a quantitative manner, the average azimuth and elevation CRLBs
\begin{align}
    \overline{\mathcal{B}}_{\varphi_0}(\alpha) &:= \frac{1}{2\pi^2} \int\limits_{\phi_0=0}^{2\pi} \int\limits_{\theta_0=0}^{\pi} \mathcal{B}_{\varphi_0}\left(\alpha, \theta_0, \varphi_0 \right) d\theta_0 d\varphi_0 \label{eq:average_CRLBs}\\
    \overline{\mathcal{B}}_{\theta_0}(\alpha) &:= \frac{1}{2\pi^2} \int\limits_{\phi=0}^{2\pi} \int\limits_{\theta=0}^{\pi} \mathcal{B}_{\theta_0}\left(\alpha, \theta_0, \varphi_0 \right) d\theta_0 d\varphi_0 \ ,\nonumber
\end{align} 
were calculated as well and summarized in Table \ref{tab:average_CRLBs} for the same three $\alpha$ values used in Figure \ref{fig:CRLBs_initial_vs_optimized}.

\begin{table}[htbp]
\centering
\caption{Antenna positions for initial and optimized antenna arrays.}
\label{tab:antenna_positions}
\setlength{\tabcolsep}{3pt}
\begin{tabular}{c|c|c}
    \hline
    & Initial Array & Optimized Array\\
    & Arrangement & Arrangement\\
    \hline $\Delta x_2$ in $\si{\milli\meter}$ & $70$ & $70$ \\
    $\Delta x_3$ in $\si{\milli\meter}$ & $0$ & $59$ \\
    $\Delta y_3$ in $\si{\milli\meter}$ & $70$ & $70$ \\
    $\beta_1$ in $\si{\degree}$ & $0$ & $67$ \\
    $\beta_2$ in $\si{\degree}$ & $0$ & $156$ \\
    $\beta_3$ in $\si{\degree}$ & $0$ & $158$ \\
    \hline
\end{tabular}
\end{table}

\begin{table}[htbp]
\centering
\caption{Average CRLBs of initial and optimized antenna arrays.} 
\label{tab:average_CRLBs}
\setlength{\tabcolsep}{3pt}
\setlength\extrarowheight{2pt}
\begin{tabular}{c|cc|cc}
    \hline
    & \multicolumn{2}{c|}{Initial Array} & \multicolumn{2}{c}{Optimized Array} \\
    $\alpha$ & $\sqrt{\overline{\mathcal{B}}_{\varphi_0}}$ & $\sqrt{\overline{\mathcal{B}}_{\theta_0}}$ & $\sqrt{\overline{\mathcal{B}}_{\varphi_0}}$ & $\sqrt{\overline{\mathcal{B}}_{\theta_0}}$ \\
    in $\si{\degree}$ & in $\si{\milli\degree} $ & in $\si{\milli\degree} $ & in $\si{\milli\degree}$ & in $\si{\milli\degree}$ \\
    \hline $0$ & $25.97$ & $43.98$ & $24.38$ & $21.10$\\
    $45$ & $14.57$ & $13.23$ & $19.21$ & $10.76$ \\
    $90$ & $15.15$ & $78.81$ & $15.1$ & $44.86$\\
    \hline
\end{tabular}
\end{table}

In the initial array, all antennas are co-oriented. As the XETS antennas are linearly polarized \cite{costaPerformanceCrossedExponentially2009}, the CRLBs $\mathcal{B}_{\varphi_0}$ and $\mathcal{B}_{\theta_0}$ are expected to strongly depend on the polarization $\alpha$ of the incident plane wave. And indeed, Figures \ref{fig:CRLBs_initial_vs_optimized}a and b show exactly this strong dependency on $\alpha$. For example, in \ref{fig:CRLBs_initial_vs_optimized}a, $\mathcal{B}_{\theta_0}$ is very low for $\alpha=0\si{\degree}$ due to the incident wave being co-polarized to the antennas. For $\alpha=90\si{\degree}$ on the other hand, $\mathcal{B}_{\theta_0}$ is larger by approximately four orders of magnitude compared to the $\alpha=0\si{\degree}$ case due to the incident wave now being cross-polarized to the antennas. The exactly opposite situation is visible in \ref{fig:CRLBs_initial_vs_optimized}b, where CRLBs for the $\varphi=90\si{\degree}$ cut are depicted. This result may look odd at first hand, but is indeed expected, considering that the unit vectors in azimuth and elevation directions switch places between the $\varphi=0\si{\degree}$ and $\varphi=90\si{\degree}$ principal cuts \cite{9714428}. As such, incident wave's polarization vector in cartesian coordinates for $\alpha=90\si{\degree}$ in the $\varphi=90\si{\degree}$ cut corresponds to the same cartesian direction than the polarization vector for $\alpha=0\si{\degree}$ in the $\varphi=0\si{\degree}$ principal cut. Comparing now the CRLBs of the initial array (dashed lines) with the CRLBs of the optimized array (solid lines), one observes that the CRLBs of the optimized array are increased compared to the initial array for certain $\alpha$ and $\theta_0$. But overall, the CRLBs decreased, indicating that localization performance is sacrificed in certain parts of the parameter, to achieve an overall improvement of the localization performance. This behavior is quantified in Table \ref{tab:average_CRLBs}, comparing the average azimuth and elevation CRLBs $\overline{\mathcal{B}}_{\varphi_0}$ and $\overline{\mathcal{B}}_{\theta_0}$ according to \eqref{eq:average_CRLBs}. Compared to the initial array, all average CRLBs decrease except for $\overline{\mathcal{B}}_{\theta_0}$ in the $\alpha=45\si{\degree}$ case, which slightly increase.

What can be observed in all subplots of Figure \ref{fig:CRLBs_initial_vs_optimized} is, that most CRLB curves rapidly increase around $\theta_0=\pm 90°$. This effect can be explained by the dipole-like radiation pattern of the XETS antennas, with little to no radiation, and reception, towards the side of the antennas. A side-effect of this dipole-like radiation pattern is that a relatively good localization performance is achieved even for electromagnetic waves incident from the back side of the array, that is for $\theta_0<-90\si{\degree}$ and $\theta_0>90\si{\degree}$.
Last but not least, Figures \ref{fig:CRLBs_initial_vs_optimized}c and d show a rapid increase of $\mathcal{B}_{\varphi_0}$ around $\theta_0=0\si{\degree}$. This behavior is not related to the actual antenna performance, but rather to the selected coordinate frame, as for $\theta_0=0\si{\degree}$, all values values of $\varphi_0$ result to the same incident wave direction in cartesian coordinates. As such, no unique estimation of $\varphi_0$ is possible at $\theta_0=0\si{\degree}$. 

\begin{figure}[htbp]
    \centering
    \begin{minipage}[b]{0.75\columnwidth}
        \centering
        \includegraphics[width=\columnwidth]{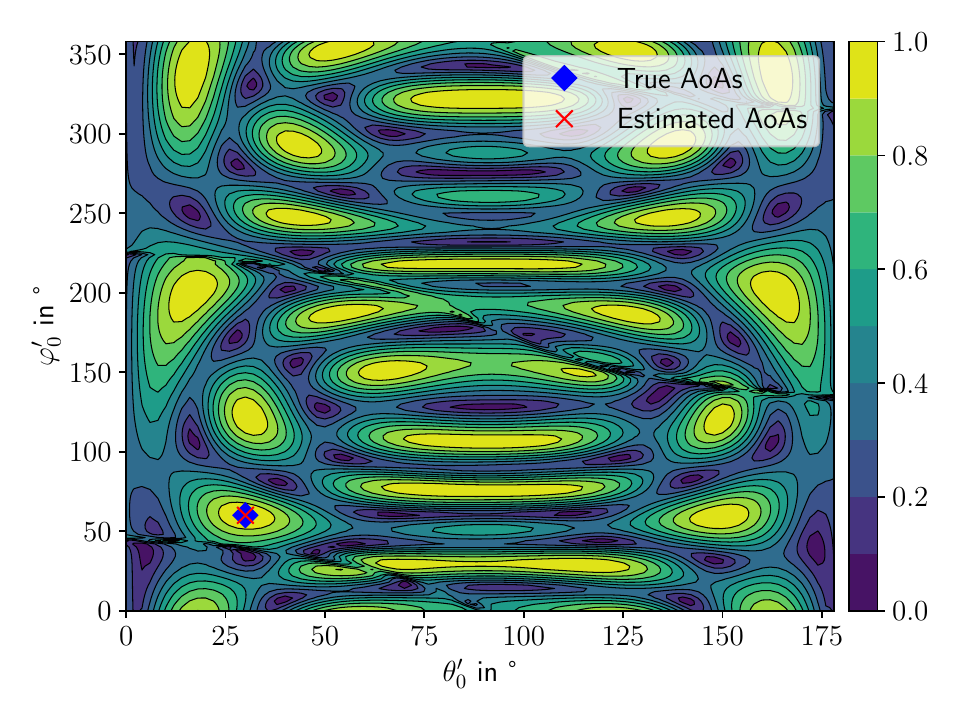}
        (a)
    \end{minipage}\\
    \begin{minipage}[b]{0.75\columnwidth}
        \centering
        \includegraphics[width=\columnwidth]{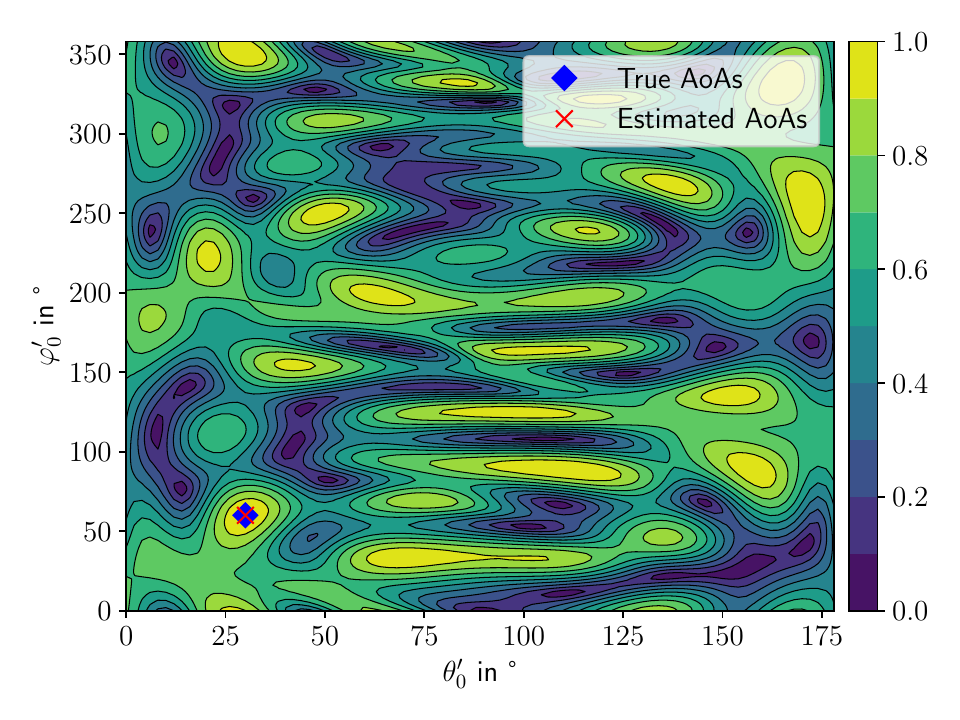}
        (b)
    \end{minipage}
    \caption{Contour plots of the normalized beampatterns $\mathcal{S}\left(\theta_{0}^\prime, \varphi_{0}^\prime \right)$ of (a) the initial array and (b) the optimal array.}
    \label{fig:beampatterns_contour_plots} 
\end{figure}

\begin{figure}[htbp]
    \centering
    \begin{minipage}[b]{0.75\columnwidth} 
        \centering
        \includegraphics[width=\columnwidth]{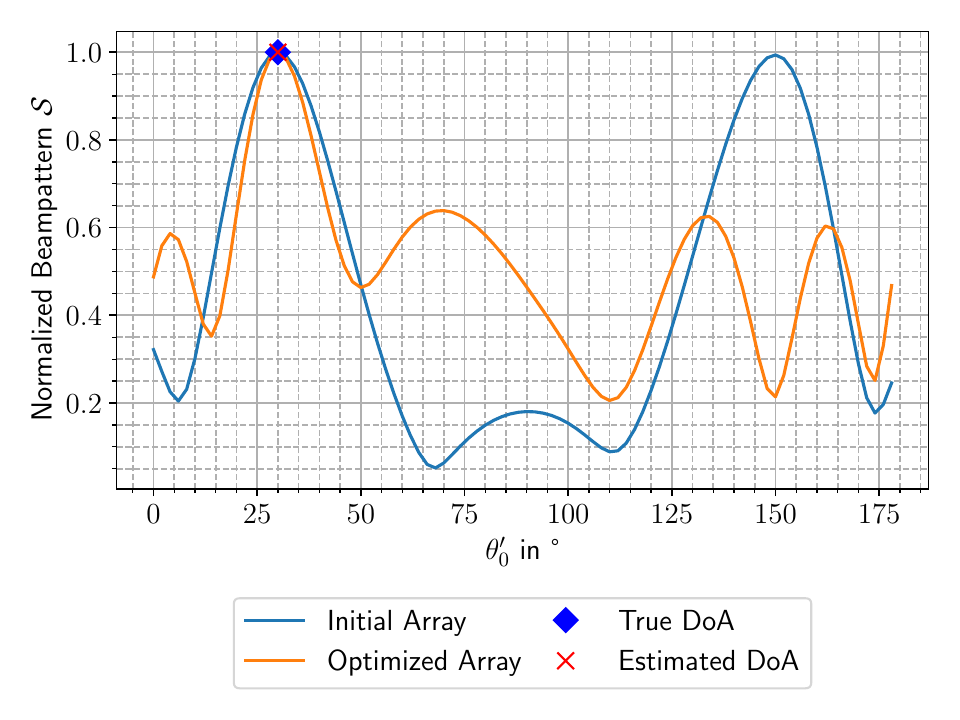} 
        (a)
    \end{minipage}\\
    \begin{minipage}[b]{0.75\columnwidth}
        \centering
        \includegraphics[width=\columnwidth]{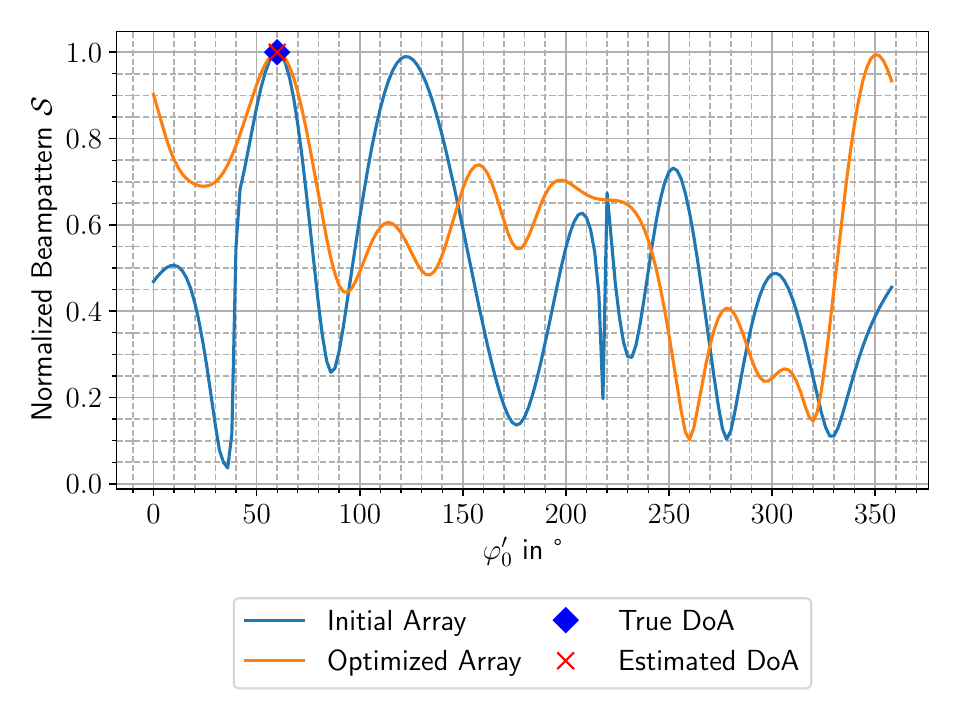}
        (b)
    \end{minipage}
    \caption{Line plots of the normalized beampatterns $S\left(\theta_{0}^\prime, \varphi_{0}^\prime \right)$. Subfigure (a) depicts the spectra over elevation $\theta_{0}^\prime$ at $\varphi_{0}^\prime=\varphi_{0}=60\si{\degree}$, whereas subfigure (b) depicts the spectra over azimuth $\varphi_{0}^\prime$ at $\theta_{0}^\prime=\theta_{0}=30\si{\degree}$.}
    \label{fig:beampatterns_lineplots} 
\end{figure}

Both, the initial and optimized array, have an antenna element spacing larger than $\frac{\lambda}{2}$ at UWB channel 9, and are thus considered as aliased arrays. 
The usage of aliased arrays for wireless localization is not necessarily a problem, as possible ambiguities can be resolved using proper localization algorithms \cite{wildingSingleAnchorMultipathAssistedIndoor2018,abramovichResolvingManifoldAmbiguities1999,gavishArrayGeometryAmbiguity1996}. However, the necessary condition for the arrays not to suffer from manifold ambiguities, derived in \eqref{eq:array_manifold_condition_final} holds for both, the initial and optimized arrays. It can thus be considered as unlikely that two array response vectors are exactly colinear. Still, one could expect several sidelobes in the array beam patterns \cite{vantreesOptimumArrayProcessing2002}. To get a better understanding of possible sidelobes in the beam patterns and thereof resulting AoA ambiguities, we calculate the normalized beam patterns for a plane wave incident from $\theta_0=30\si{\degree}$ and $\varphi_0=60\si{\degree}$ for both arrays, assuming a known propagation delay and polarization.
The normalized conventional beam pattern $\mathcal{S}$ \cite[Sec. 2.5]{vantreesOptimumArrayProcessing2002} over elevation $\theta_{0}^\prime$ azimuth $\varphi_{0}^\prime$, for a plane wave incident from elevation $\theta_0$ and azimuth $\varphi_0$ is calculated as
\begin{align}
   \mathcal{S}\left(\theta_{0}^\prime, \varphi_{0}^\prime \right) = \frac{\|\myVec{a}^H(\theta_{0}^\prime, \varphi_{0}^\prime) \myVec{a}(\theta_{0}, \varphi_{0})\|}{\|\myVec{a}(\theta_{0}^\prime, \varphi_{0}^\prime)\| \|\myVec{a}(\theta_{0}, \varphi_{0})\|} \label{eq:beam_pattern}
\end{align}
with $\myVec{a}$ from \eqref{eq:array_manifold}.
Relation \eqref{eq:beam_pattern} is evaluated for $\theta_{0}^\prime\in\left[0, \pi \right]$ and $\theta_{0}^\prime\in\left[0, 2\pi \right]$ for both, the initial and optimized antenna arrays. The polarization of the incident plane wave considered during the evaluation was $\alpha=45\si{\degree}$. 
Contour and line plots of the normalized beam patterns are shown in Figures \ref{fig:beampatterns_contour_plots} and \ref{fig:beampatterns_lineplots}. The figures reveal several sidelobes being present in the beam patterns, resulting from the antenna spacings being $>\frac{\lambda}{2}$. Nevertheless, the absolute maxima of the beam patterns occur at the true AoAs, also marked in Figures \ref{fig:beampatterns_contour_plots} and \ref{fig:beampatterns_lineplots}. Several other sidelobes are visible in the beam patterns, especially in Figure \ref{fig:beampatterns_lineplots}, appearing to have almost the same height as the mainlobe. Numerically assessing the heights of the largest sidelobes reveals heights of close to $0.99$. As such, there are no manifold ambiguities present in the exact sense, namely that two array responses are exactly colinear.

\section{Conclusion \& Outlook}
\label{sec:conclusion_and_outlook}
We proposed an optimization procedure for determining the positions and orientations of antenna elements in sparse arrays for improving the localization performance. In contrast to to state-of-the-art methods, our proposed method considers the full frequency-, direction- and polarization-dependent reception characteristics of the antenna array, including mutual coupling.
The procedure has been demonstrated by means of a 3-element XETS array under consideration of linearly polarized incident waves and mutual coupling between the antenna elements.
The optimization started with an array configuration that provides good performance for incident waves co-polarized to the antenna elements, but drops in performance for other polarizations. 
By moving and rotating the individual antenna elements, the optimization procedure achieved a lower variation of the localization performance for different polarizations compared to the initial antenna array.\\
The main limitations of our proposed optimization procedure, are its high computational demands and extended execution time. The source of both limitations is the full-wave EM simulation executed for
each antenna arrangement investigated during the optimization. A reduction of computation time could be achieved by, for example, implementing one of the approaches from either \cite{rubioGeneralizedscatteringmatrixAnalysisClass2005} or \cite{marinovicFastCharacterizationMutually2021}.\\
The proposed optimization procedure targets the optimization of antenna arrays for determining distance and direction of a single transmitter relative to the agent under AWGN channel conditions. But if the agent operates under less relaxed channel conditions, e.g. if the wireless channel suffers from multipath propagation, the delivered arrays may be suboptimal, in the sense that they do not allow a proper resolution of all multipath components. Considering such channel conditions in the optimization procedure requires the adaption of the underlying signal models, such that the resulting FIM reflects the ability of the arrays to resolve multipath components \cite{SheWin:TIT2010part1,LeitingerJSAC2015,wildingSingleAnchorMultipathAssistedIndoor2018}.  \\
Investigations on resolving the limitations of the proposed optimization procedure in terms of execution time and consideration of multipath propagation, are considered to be subject of future research activities.

% \section*{Acknowledgment} 
% The work presented in this document was conducted in the frame of the SINFONIA project. The SINFONIA has received funding from the Recovery and Resilience Facility (RRF) as the centrepiece of NextGenerationEU via the Austrian Research Promotion Agency (FFG) and Austria Wirtschaftsservice Gesellschaft mbH (aws) in the frame of the IPCEI ME/CT – Important Project of Common European Interest on Microelectronic and Communication Technologies under FFG project No 917423 and AWS project No P2431566.

\appendices

\section{Wideband Plane Wave Model}
\label{sec:wideband_plane_wave_model}
\begin{figure}[!tb]
    \centering
    \includegraphics[width=\columnwidth]{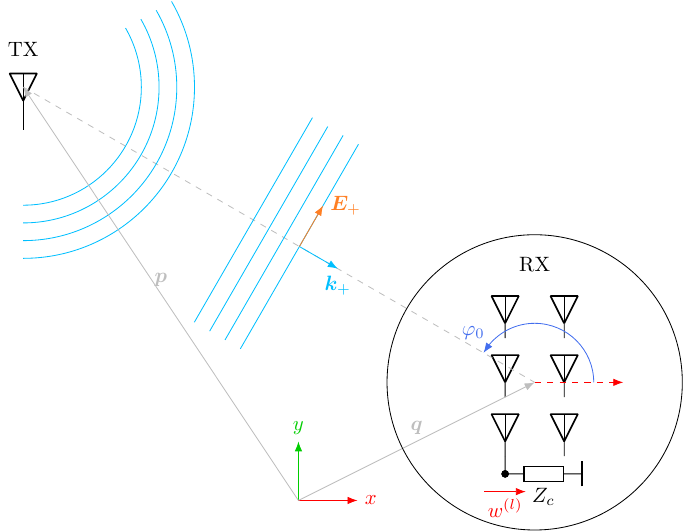}
    \caption{A sketch of a plane wave with propagation vector $\myVec{k}_+$ in incident on antenna array. The relations between the electric field strength $\myVec{E}_+$ and the resulting outward-propagating voltage waves $w^{(l)}$ are efficiently described by VSWFs.}
    \label{fig:plane_wave_model}
\end{figure} 
The SWCs of a wideband plane wave incident on an AUT, are derived by comparison of coefficients with the VSWE of narrowband plane waves found in the literature \cite{hansenSphericalNearfieldAntenna1988a,strattonElectromagneticTheory1941}.
The plane wave model used in this section is a simplified version of the model presented in \cite{laferEffectiveHeightAnalysisUWB2025}. 
The transmitter \textit{TX} in Figure \ref{fig:plane_wave_model} is assumed be an isotropic radiator with infinite bandwidth and direction-independent polarization $\myVec{P}$.
The excitation signal of the transmitter is a real-valued lowpass signal $s(t)$ with spectrum $S(\omega)$, up-converted to the desired RF channel with angular carrier frequency $\omega_0$. 
The emitted electromagnetic wave propagates in free space, and arrives at AUT as a plane wave with propagation vector $\myVec{k}_+$ and electric field strength $\myVec{E}_+$. The passband representation of $\myVec{E}^+$ in time-domain is
\begin{align}
    \myVec{e}_+(t) = \frac{A\myVec{P}}{r_0} \sqrt{\frac{Z_0}{Z_c}} \frac{d}{dt} \left(s(t) e^{i\omega_0 t} \right) * \delta\left(t - \myVec{k}_+^T \myVec{q} - \tau\right)\ , \label{eq:e_incident_time_domain}
\end{align}
with propagation vector $\myVec{k}_+ = \frac{\omega}{c_0}\myVec{n}_+$, propagation delay $\tau = \frac{r_0}{c_0}$ where $r_0$ is the euclidean distance between the positions of the transmitter $\myVec{p}\in\mathbb{R}^3$ and the position of the AUT $\myVec{q}\in\mathbb{R}^3$, and $A\in\mathbb{R}$ being a proportionality factor modelling the signal amplitude. Vector $\myVec{n}_+\in\mathbb{R}^3$ is a unit vector denoting the propagation direction of the plane wave towards the AUT. 
The polarization $\myVec{P}$ in spherical coordinates is
\begin{align*}
    \myVec{P} \overset{\triangle}{=} \myVec{P}\left(P_\theta, \phi_\theta, P_\varphi, \phi_\varphi\right) = P_\theta e^{i\phi_\theta} \myVec{i}_{\theta} + P_\varphi e^{i\phi_\varphi} \myVec{i}_{\varphi}\ .
\end{align*}
The frequency domain representation of \eqref{eq:e_incident_time_domain} is
\begin{align}
    \myVec{E}_+(\omega) = A\sqrt{\frac{Z_0}{Z_c}} \frac{i\omega}{2\pi c_0 r_0} \myVec{P} S(\omega - \omega_0) e^{-i \myVec{k}_+^T \myVec{q}} e^{-i\omega \tau} \ .  \label{eq:e_incident_frequency_domain}
\end{align}
The spectrum \eqref{eq:e_incident_frequency_domain} is a passband spectrum with non-zero frequency components within $\omega_0-B\leq \omega \leq \omega_0+B$, with $B$ being the low-pass bandwidth of $s(t)$ in $\si{\radian\per\second}$.\newline

According to \cite{hansenSphericalNearfieldAntenna1988a}, the SWCs of a plane wave with propagation vector $\myVec{k}_0$ and electric field strength
\begin{align}
    \myVec{E} = \myVec{E}_0 e^{i \myVec{k}_0^T \myVec{q}} \label{eq:plane_wave_hansen} 
\end{align}
are obtained as
\begin{align}
    a_{smn}=b_{smn}=\frac{\sqrt{\eta}}{2k}(-1)^m \sqrt{4\pi}i \myVec{E}_0^T \myVec{K}_{s(-m)m}\ , \label{eq:SWCs_plane_wave_hansen}
\end{align}
where $\myVec{K}_{smn}$ are the \textit{far-field pattern functions} given in \eqref{eq:far_field_pattern_functions}. 

The SWCs of \eqref{eq:e_incident_frequency_domain} cannot be determined through direct comparison with \eqref{eq:plane_wave_hansen} and \eqref{eq:SWCs_plane_wave_hansen}, due to the different underlying time- and distance-dependencies assumed by the VSWFs \cite{IEEERecommendedPractice2012,soklicFullSphereAntennaMeasurements2024}.
The VSWFs \eqref{eq:VSWE_general} assume an $e^{-i\omega t}$ time dependence and an $e^{ikr}$ distance dependence \cite{hansenSphericalNearfieldAntenna1988a,IEEERecommendedPractice2012,strattonElectromagneticTheory1941}, in contrast to the $e^{i\omega t}$ time- and $e^{-ikr}$ distance dependencies of \eqref{eq:e_incident_time_domain} and \eqref{eq:e_incident_frequency_domain}. This issue is commonly addressed by applying a complex conjugation to the respective time-domain signals. Utilizing the Fourier transform property
\begin{align*}
    x(t) \overset{\scriptscriptstyle\mathcal{F}}{\Leftrightarrow} X(\omega) \Rightarrow x^*(t) \overset{\scriptscriptstyle \mathcal{F}}{\Leftrightarrow} X^*(-\omega)\ ,
\end{align*}
one obtains
\begin{flalign*}
    &\myVec{\hat{e}}_+(t) := \left\{\myVec{e}^+(t)\right\}^* \Rightarrow& \\[0.5em]
    &\hat{\myVec{E}}_+(\omega) = \mathcal{F}\left\{ \myVec{\hat{e}}_+(t) \right\} = \left\{\myVec{E}_+(-\omega)\right\}^* = &\\
    &\quad = A\sqrt{\frac{Z_0}{Z_c}} \frac{i\omega}{2\pi c_0 r_0} \myVec{P}^* S^*(-\omega - \omega_0) e^{i\myVec{k}_+^T \myVec{q}}\ e^{-i\omega \tau} \ .&
\end{flalign*}
Considering that $S(\omega)$ is Hermitian symmetric due to $s(t)$ being a real-valued signal, one finally obtains the passband representation of the electric field strength of the incident plane wave in the correct time basis as
\begin{align}
    \hat{\myVec{E}}_+(\omega) \label{eq:plane_wave_RF} = A\sqrt{\frac{Z_0}{Z_c}} \frac{i\omega}{2\pi c_0 r_0} \myVec{P}^* S(\omega + \omega_0) e^{i\myVec{k}_+^T \myVec{q}}\ e^{-i\omega \tau} \ . 
\end{align}
Relation \eqref{eq:plane_wave_RF} also represents a passband spectrum, but this time with non-zero frequency components only within $-\omega_0-B \leq \omega\leq -\omega_0+B$.
By comparing \eqref{eq:plane_wave_RF} with \eqref{eq:plane_wave_hansen}, one obtains complex amplitude $\myVec{E}_0$ of the incident plane wave \eqref{eq:plane_wave_RF}
\begin{align}
    \myVec{E}_0 = A\sqrt{\frac{Z_0}{Z_c}} \frac{i\omega}{2\pi c_0 r_0} \myVec{P}^* S(\omega + \omega_0) e^{-i\omega\tau}\ . \label{eq:plane_wave_complex_amplitude}
\end{align}
Subsequent comparison of \eqref{eq:plane_wave_complex_amplitude} with \eqref{eq:SWCs_plane_wave_hansen} delivers the desired wideband SWCs for the plane wave \eqref{eq:plane_wave_RF}
\begin{flalign}
    & a_{smn} = b_{smn} = \frac{\sqrt{\eta}}{2k}(-1)^m \sqrt{4\pi}i \bigg[ \label{eq:wideband_plane_wave_SWCs_tmp} & \\
    & \quad A\sqrt{\frac{Z_0}{Z_c}} \frac{i\omega}{2\pi c_0 r_0} \myVec{P}^* S(\omega - \omega_0) e^{-i\omega\tau} \bigg]^T \myVec{K}_{s(-m)n} \nonumber&
\end{flalign}
One can simplify \eqref{eq:wideband_plane_wave_SWCs_tmp} by considering that $k=\frac{\omega}{c_0}$ and $\eta = \frac{1}{Z_0}$:
\begin{align}
    a_{smn} = b_{smn} = -A \frac{(-1)^m e^{-i\omega\tau}}{2r_0\sqrt{\pi Z_c}}  S(\omega - \omega_0) \myVec{P}^H \myVec{K}_{s(-m)n}\ . \label{eq:wideband_plane_wave_SWCs_passband}
\end{align}

\section{Matrix and Vector Definitions for the Discrete Frequency Model}
\label{sec:matrix_vector_definitions}

The frequency-domain baseband representations of the antenna port voltages $w^{(l)}$ according to \eqref{eq:single_antenna_RX_wideband},
\begin{flalign}
    & \tilde{w}^{(l)}(\omega, \theta_0, \varphi_0) = \sum\limits_{smn} \bigg[ \tilde{R}_{smn}^{(l)}(\omega) \frac{-A(-1)^m e^{-i(\omega+\omega_0)\tau}}{2r_0\sqrt{\pi Z_c}} \label{eq:bb_port_voltages_appendix}& \\
    &\quad S(\omega) \myVec{P}^T \myVec{K}_{s(-m)n}^* (\theta_0, \varphi_0) \bigg]\nonumber \ ,
\end{flalign}
are sampled in frequency domain at the discrete frequencies $\omega_p = p\Delta \omega$ with $\Delta \omega = \frac{2\pi f_s}{P}$, $p\in\left[\frac{-P+1}{2}, \frac{P-1}{2}\right]$ and $P$ being an odd integer number.
The resulting $P$ discrete-frequency samples
\begin{align*}
    \tilde{w}^{(l)}[p] = w^{(l)}(p\Delta \omega)\ ,
\end{align*}
of all $L$ antennas are collected in a large vector $\myVec{\tilde{w}}$. Using the definitions from \eqref{eq:matrix_vector_definitions} below, $\myVec{\tilde{w}}$ can be expressed in matrix-vector notation by
\begin{flalign*}
    &\myVec{\tilde{w}} = \left(\myVec{1}_{L\times 1} \otimes \myVec{\tau}(\tau)\right) \odot \left(\myVec{R} \myVec{M}\myVec{K}^H(\theta_0, \varphi_0) \myVec{P}\right) \odot \left(\myVec{1}_{L\times 1} \otimes \myVec{S}\right)\ ,&
\end{flalign*}
where $\myVec{1}_{N\times M}$ is a real-valued matrix of dimension $N\times M$ filled with ones. Details regarding the matrix definitions are provided below, but require some clarifications regarding the triple and single-index conventions of spherical modes.

The joined mode index $j$ can be calculated from the original mode index triplet $(s,m,n)$ by
\begin{align}
j[s,m,n] = 2\left(n(n+1) + m-1\right) + s \label{eq:single_index_from_triple_index_appendix}
\end{align} 
This operation is invertible \cite{hansenSphericalNearfieldAntenna1988a}, and one obtains unique $s,m,n$ for each $j$ via 
\begin{align}
    s[j]&=\begin{cases}
        1, & j=\text{odd}\\
        2, & j=\text{even}
    \end{cases} \nonumber \\
    n[j] &= \text{integer part of}\ \sqrt{\frac{j-s[j]}{2}+1} \nonumber \\
    m[j] &= \frac{j-s[j]}{2}+1-n[j]\left(n[j]+1\right) \label{eq:triple_index_from_single_index}
\end{align}
Another index $j_K$ is defined, allowing to index the far-field functions $\myVec{K}_{s(-m)n}$ in \eqref{eq:bb_port_voltages_appendix} also by single index, despite the negative sign of $m$:
\begin{align}
    \hat{j}[s,m,n] := 2\left(n(n+1) - m-1\right) + s\ . \label{eq:single_index_negative_m}
\end{align}  
Combining \eqref{eq:single_index_from_triple_index_appendix}, \eqref{eq:triple_index_from_single_index} and \eqref{eq:single_index_negative_m} allows to express $j_K$ in terms of $j$ by
\begin{align*}
    \hat{j}[j] = 2\left(n[j](n[j]+1) - m[j]-1\right) + s[j]= j - 4m[j]\ .
\end{align*}
The final matrix and vector definitions are then:
\begin{flalign}
    &\myVec{S} := \frac{-A}{2r_0\sqrt{\pi Z_c}}\left[S\left(\frac{-P+1}{2}\Delta \omega\right), \hdots,  S\left(\frac{P-1}{2}\Delta \omega\right)\right]^T\nonumber &\\[1em]
    &\myVec{\tau}(\tau) := \left[ e^{-i\left(\frac{-P+1}{2}\Delta \omega+\omega_0\right)\tau}, \hdots, e^{-i\left(\frac{P-1}{2}\Delta \omega+\omega_0\right)\tau} \right]^T \nonumber&\\[1em]
    &\tilde{R}_{j,p}^{(l)} := \tilde{R}_j^{(l)}(p\Delta \omega)  \nonumber&\\[0.5em]
    &\myVec{\tilde{R}}^{(l)} := \begin{bmatrix}
        \tilde{R}_{1,\frac{-P+1}{2}}^{(l)} & \hdots & \tilde{R}_{J,\frac{-P+1}{2}}^{(l)} \\
        \vdots & & \vdots \\
        \tilde{\tilde{R}}_{1,\frac{P-1}{2}}^{(l)} & \hdots & \tilde{R}_{J,\frac{P-1}{2}}^{(l)}
    \end{bmatrix}, \quad \myVec{R} := \begin{bmatrix}
        \myVec{\tilde{R}}^{(1)} \\ \vdots \\ \myVec{\tilde{R}}^{(L)} \nonumber
    \end{bmatrix} \nonumber&\\[1em]
    &\myVec{M} := \text{diag}\left\{ (-1)^{m[1]}, \hdots, (-1)^{m[J]} \right\} \nonumber &\\[1em]
    &\myVec{K} := \left[\myVec{K}_{\hat{j}[1]}, \hdots,  \myVec{K}_{\hat{j}[J]} \right]^T \label{eq:matrix_vector_definitions}
\end{flalign}

\section{Type-I manifold ambiguities}
\label{sec:manifold_ambiguities}
Defining the array manifold 
\begin{align}
    \myVec{a}\left( \myVec{\theta}\right) := \left(\myVec{1}_{L\times 1} \otimes \myVec{\tau}(\tau)\right) \odot \left(\myVec{R} \myVec{M}\myVec{K}^T(\theta_0, \varphi_0) \myVec{P}^*\right)\ , \label{eq:array_manifold}
\end{align}
and considering two distinct parameters vectors $\myVec{\theta}_1 \neq \myVec{\theta}_2$ and a non-zero complex scalar $\mathcal{K}$, a necessary condition for
\begin{align}
    \myVec{a}\left( \myVec{\theta}_1\right) - \mathcal{K} \myVec{a}\left( \myVec{\theta}_2\right) \neq \myVec{0} \label{eq:array_manifold_condition_tmp}
\end{align}
is derived in this section. Inserting \eqref{eq:array_manifold} in \eqref{eq:array_manifold_condition_tmp} delivers
\begin{flalign*}
    & \myVec{a}\left( \myVec{\theta}_1\right) - \mathcal{K} \myVec{a}\left( \myVec{\theta}_2\right) = \left( \myVec{1}_{L\times 1} \otimes \left[\myVec{\tau}(\tau_1) - \mathcal{K}\myVec{\tau}(\tau_2)  \right] \right) \odot \myVec{R} \myVec{M} \bigg( &\\
    & \quad \myVec{K}^T(\theta_{1}, \varphi_1) \myVec{P}_1^* - \mathcal{K} \myVec{K}^T(\theta_{2}, \varphi_2) \myVec{P}_2^*\bigg) &
\end{flalign*}
For notational convenience, we summarize the terms
\begin{align}
    \myVec{u}(\myVec{\theta}_1, \myVec{\theta}_2) := \myVec{K}^T(\theta_{1}, \varphi_1) \myVec{P}_1^* - \mathcal{K} \myVec{K}^T(\theta_{2}, \varphi_2) \myVec{P}_2^*
\end{align}
in \eqref{eq:array_manifold_condition_tmp}, which delivers
\begin{flalign}
    &\myVec{a}\left( \myVec{\theta}_1\right) - \mathcal{K} \myVec{a}\left( \myVec{\theta}_2\right) = \left( \myVec{1}_{L\times 1} \otimes \left[\myVec{\tau}(\tau_1) - \mathcal{K}\myVec{\tau}(\tau_2)  \right] \right) \odot & \nonumber \\
    &\quad \myVec{R} \myVec{M} \myVec{u}(\myVec{\theta}_1, \myVec{\theta}_2)\neq \myVec{0}\ .& \label{eq:array_manifold_condition_final}
\end{flalign}

Considering the matrix-vector definitions \eqref{eq:matrix_vector_definitions}, the first term in \eqref{eq:array_manifold_condition_final},
\begin{align*}
    \myVec{1}_{L\times 1} \otimes \left[\myVec{\tau}(\tau_1) - \mathcal{K}\myVec{\tau}(\tau_2)\right]\ ,
\end{align*}
is always $\neq \myVec{0}$ if $\tau_1 \neq \tau_2$, provided that both propagation delays fulfill the unambiguous delay condition \eqref{eq:unambiguous_delay_condition}.
Thus, a necessary (but not sufficient) condition for \eqref{eq:array_manifold_condition_final} to hold true, is that
\begin{align}
    \myVec{R} \myVec{M} \myVec{u}(\myVec{\theta}_1, \myVec{\theta}_2) = \myVec{0}\ ,
\end{align}
has only the trivial solution $\myVec{u}=\myVec{0}$. This holds true if matrix $\myVec{R} \myVec{M}$ has full rank. But as matrix $\myVec{M}$ is a diagonal matrix with non-zero diagonal elements according to \eqref{eq:matrix_vector_definitions}, the actual necessary condition for \eqref{eq:array_manifold_condition_final} hold true is to verify that the matrix of array reception coefficients $\myVec{R}$ has full rank.

\section{Partial Derivatives required for the Fisher Information Matrix}
\label{sec:FIM_partial_derivatives}
This appendix covers the partial derivatives of \eqref{eq:AWGN_signal_vector_likelihood} with respect to the signal parameters \eqref{eq:parameter_vector}, required for calculating the FIM \eqref{eq:FIM}.
Fortunately, six out of the seven partial derivatives are straight-forward, only the partial derivative with respect to $\myVec{\theta}_0$ requires more treatment.

Considering the matrix-vector definitions from \eqref{eq:matrix_vector_definitions}, one obtains for the partial derivative with respect to the propagation delay $\tau$
\begin{align}
    \frac{\partial \myVec{\tilde{w}}}{\partial \tau} = \left[\myVec{1}_{L\times 1}\otimes -i(\Delta \omega \myVec{p}-\omega_0) \right] \odot \myVec{\tilde{w}}\ ,
\end{align}
with $\myVec{p} := \left[\frac{-P-1}{2}, \hdots, \frac{P-1}{2} \right]$.
The next four partial derivatives are with respect to the four polarization parameters $P_\theta$, $P_\varphi$, $\phi_\theta$ and $\phi_\varphi$. Under consideration of \eqref{eq:polarization_vector} one obtains
\begin{flalign}
    &\frac{\partial \myVec{\tilde{w}}}{\partial P_\theta} = \left(\myVec{1}_{L\times 1} \otimes \myVec{\tau}\right) \odot e^{-i\phi_\theta }\left(\myVec{R} \myVec{M}\myVec{K}^T \myVec{i}_\theta\right) \odot \left(\myVec{1}_{L\times 1} \otimes \myVec{S}\right) \nonumber & \\
    &\frac{\partial \myVec{\tilde{w}}}{\partial P_\varphi} = \left(\myVec{1}_{L\times 1} \otimes \myVec{\tau}\right) \odot e^{-i\phi_\varphi }\left(\myVec{R} \myVec{M}\myVec{K}^T \myVec{i}_\varphi\right) \odot \left(\myVec{1}_{L\times 1} \otimes \myVec{S}\right) \nonumber &\\
    &\frac{\partial \myVec{\tilde{w}}}{\partial \phi_\theta} = \left(\myVec{1}_{L\times 1} \otimes \myVec{\tau}\right) \odot (-i)P_\theta e^{-i\phi_\theta }\left(\myVec{R} \myVec{M}\myVec{K}^T \myVec{i}_\theta\right) \odot  \nonumber &\\
    &\quad \odot \left(\myVec{1}_{L\times 1} \otimes \myVec{S}\right) \nonumber &\\
    &\frac{\partial \myVec{\tilde{w}}}{\partial \phi_\varphi} = \left(\myVec{1}_{L\times 1} \otimes \myVec{\tau}\right) \odot (-i)P_\varphi e^{-i\phi_\varphi }\left(\myVec{R} \myVec{M}\myVec{K}^T \myVec{i}_\varphi\right) \odot & \nonumber \\
    & \quad \odot \left(\myVec{1}_{L\times 1} \otimes \myVec{S}\right) \quad &
\end{flalign}
with $\myVec{i}_\theta$ and $\myVec{i}_\varphi$ being the spatial unit vectors in $\theta$ and $\varphi$ direction in spherical coordinates. 
The remaining derivatives are with respect to the incident wave direction expressed by $\theta_0$ and $\varphi_0$, only occurring in the matrix of far-field pattern functions $\myVec{K}$ in \eqref{eq:single_antenna_RX_wideband}. As such, we require
\begin{align}
    \frac{\partial \myVec{K}}{\partial \theta_0} = \left[\frac{\partial \myVec{K}_{j_{\myVec{K}}[1]}}{\partial \theta_0} , \hdots,  \frac{\partial \myVec{K}_{j_{\myVec{K}}[J]}}{\partial \theta_0} \right]^T\ , \label{eq:ff_matrix_derivatives}\\[1em]
    \frac{\partial \myVec{K}}{\partial \varphi_0} = \left[\frac{\partial \myVec{K}_{j_{\myVec{K}}[1]}}{\partial \varphi_0} , \hdots,  \frac{\partial \myVec{K}_{j_{\myVec{K}}[J]}}{\partial \varphi_0} \right]^T\ . \nonumber
\end{align}

The far-field pattern functions are according to \cite{hansenSphericalNearfieldAntenna1988a}
\begin{flalign}
    & \myVec{K}_{1mn} = \sqrt{\frac{2}{n(n+1)}}\left(-\frac{m}{|m|}\right)^m e^{im\varphi_0} (-i)^{n+1}\bigg\{ &\nonumber \\
    & \quad \frac{im \bar{P}_{n}^{|m|}(\cos \theta_0)}{\sin \theta_0} \myVec{i}_\theta - \frac{d \bar{P}_{n}^{|m|}(\cos \theta_0)}{d\theta_0} \myVec{i}_\varphi \bigg\} \ , & \nonumber \\[1em]
    & \myVec{K}_{2mn} = \sqrt{\frac{2}{n(n+1)}}\left(-\frac{m}{|m|}\right)^m e^{im\varphi_0} (-i)^{n}\bigg\{ & \nonumber\\
    & \quad \frac{d \bar{P}_{n}^{|m|}(\cos \theta_0)}{d\theta_0} \myVec{i}_\theta + \frac{im \bar{P}_{n}^{|m|}(\cos \theta_0)}{\sin \theta_0} \myVec{i}_\varphi \bigg\}\ . & \label{eq:Ksmn}
\end{flalign}
Parameter $\varphi_0$ in \eqref{eq:Ksmn} only occurs in $e^{im\varphi_0}$, which leads to the partial derivatives
\begin{align}
    \frac{\partial \myVec{K}_{smn}}{\partial \varphi_0} =  im \myVec{K}_{smn} \ . \label{eq:ff_functions_d_phi}
\end{align}
The desired derivative of \eqref{eq:AWGN_signal_vector_likelihood} with respect to $\varphi_0$ is thus obtained as
\begin{flalign*}
    &\frac{\partial \myVec{\tilde{w}}}{\partial \varphi_0} = \left(\myVec{1}_{L\times 1} \otimes \myVec{\tau}\right) \odot \left(\myVec{R} \myVec{M}\frac{\partial \myVec{K}^T}{\partial \varphi_0} \myVec{P}^*\right) \odot  \left(\myVec{1}_{L\times 1} \otimes \myVec{S}\right) \nonumber \ ,&\\
\end{flalign*}
with $\frac{\partial \myVec{K}^T}{\partial \varphi_0}$ according to \eqref{eq:ff_matrix_derivatives} and \eqref{eq:ff_functions_d_phi}.
The remaining derivative of the far-field patterns functions $\myVec{K}_{smn}$ with respect to $\theta_0$, require the derivatives
\begin{align}
    \frac{d}{\theta_0} \frac{im \bar{P}_{n}^{|m|}(\cos \theta_0)}{\sin \theta_0}\ \text{and}\ \frac{d^2 \bar{P}_{n}^{|m|}(\cos \theta_0)}{d\theta_0^2}\ .\label{eq:legendre_polynomials_derivatives}
\end{align}
For notational convenience, we denote in the following
\begin{align*}
    P_n^{\prime|m|} := \frac{d \bar{P}_{n}^{|m|}(\cos \theta_0)}{d\theta_0}\ ,
\end{align*}
and drop the explicit $\cos\theta_0$ dependency of the associated Legendre polynomials $\bar{P}_{n}^{|m|}$.
The first derivate is obtained by applying the quotient rule 
\begin{align}
    \frac{d}{\theta_0} \frac{im \bar{P}_{n}^{|m|}}{\sin \theta_0} = im\frac{P_n^{\prime|m|} \sin\theta_0 - \bar{P}_{n}^{|m|} \cos\theta_0}{\sin^2\theta_0}\ .\label{eq:dPmnDivSinTheta_dTheta}
\end{align}
Considering the recurrence relation \cite{hansenSphericalNearfieldAntenna1988a}
\begin{flalign*}
    & \frac{d \bar{P}_{n}^{|m|}}{d\theta_0} = &\\
    & \quad = \begin{cases}
        -\bar{P}_n^1, & m=0 \\
        \frac{1}{2}\left[ (n-|m|+1)(n+|m|) \bar{P}_{n}^{|m|+1} -  \bar{P}_{n}^{|m|} \right] & |m| > 0
    \end{cases} \nonumber\ ,&
\end{flalign*}
the second derivative in \eqref{eq:legendre_polynomials_derivatives} can be calculated by
\begin{flalign}
    &\frac{d^2 \bar{P}_{n}^{|m|}}{d\theta_0^2} := \frac{d \bar{P}_{n}^{\prime|m|}}{d\theta_0} = \label{eq:d2Pnm_dTheta2} &\\
    & \quad = \begin{cases}
        -\bar{P}_n^{\prime 1}, & m=0 \\
        \frac{1}{2}\left[ (n-|m|+1)(n+|m|) \bar{P}_{n}^{\prime|m|+1} -  \bar{P}_{n}^{\prime|m|} \right] & |m| > 0
    \end{cases} \nonumber\ .& 
\end{flalign}
One finally obtains 
\begin{flalign}
    & \frac{\partial \myVec{K}_{1mn}}{\partial \theta_0} = \sqrt{\frac{2}{n(n+1)}}\left(-\frac{m}{|m|}\right)^m e^{im\varphi_0} (-i)^{n+1}\bigg\{ &\nonumber \\
    & \quad \frac{d}{\theta_0} \frac{im \bar{P}_{n}^{|m|}}{\sin \theta_0} \myVec{i}_\theta - \frac{d^2 \bar{P}_{n}^{|m|}}{d\theta_0^2} \myVec{i}_\varphi \bigg\}\ , & \nonumber \\[1em]
    & \frac{\partial \myVec{K}_{2mn}}{\partial \theta_0} = \sqrt{\frac{2}{n(n+1)}}\left(-\frac{m}{|m|}\right)^m e^{im\varphi_0} (-i)^{n}\bigg\{ & \nonumber\\
    & \quad \frac{d^2 \bar{P}_{n}^{|m|}}{d\theta_0^2} \myVec{i}_\theta + \frac{d}{\theta_0} \frac{im \bar{P}_{n}^{|m|}}{\sin \theta_0} \myVec{i}_\varphi \bigg\}\ ,& \label{eq:dKsmn_dTheta0}
\end{flalign}
with $\frac{d}{d\theta_0}\frac{im \bar{P}_{n}^{|m|}}{\sin \theta_0}$ and $\frac{d^2 \bar{P}_{n}^{|m|}}{d\theta_0^2}$ according to \eqref{eq:dPmnDivSinTheta_dTheta} and \eqref{eq:d2Pnm_dTheta2}.
The derivative of \eqref{eq:AWGN_signal_vector_likelihood} with respect to $\theta_0$ is thus obtained as
\begin{flalign*}
    &\frac{\partial \myVec{\tilde{w}}}{\partial \theta_0} = \left(\myVec{1}_{L\times 1} \otimes \myVec{\tau}\right) \odot \left(\myVec{R} \myVec{M}\frac{\partial \myVec{K}^T}{\partial \theta_0} \myVec{P}^*\right) \odot  \left(\myVec{1}_{L\times 1} \otimes \myVec{S}\right) \nonumber \ ,&
\end{flalign*}
with $\frac{\partial \myVec{K}^T}{\partial \theta_0}$ according to \eqref{eq:ff_matrix_derivatives} and \eqref{eq:dKsmn_dTheta0}.

\bibliographystyle{IEEEtranDOI}
\bibliography{IEEEabrv,bibliography,IEEEbibControl}

\end{document}